\documentclass[pdflatex,sn-nature]{sn-jnl}% Basic Springer Nature Reference Style/Chemistry Reference Style

\usepackage{hyperref}[citecolor=magenta]

\hypersetup{
    colorlinks=true,
    citecolor=blue,
    linkcolor=blue,
    urlcolor=blue
}

\usepackage{url}            % simple URL typesetting
\usepackage{booktabs}       % professional-quality tables
\usepackage{amsfonts}       % blackboard math symbols
\usepackage{nicefrac}       % compact symbols for 1/2, etc.
\usepackage{microtype}      % microtypography
\usepackage{xcolor}         % colors

\usepackage{enumitem}       % [leftmargin=20pt]
\usepackage{multirow}       % 
\usepackage{xspace}         % xspace
\usepackage{wrapfig}        % wrapfigure
\usepackage{makecell}       % divide row in table
\usepackage{colortbl}
\usepackage{bbm}            % indicator function (\mathbbm)
\usepackage{adjustbox}

\usepackage[most,skins,theorems]{tcolorbox}
\tcbset{
  aibox/.style={
    width=\linewidth,
    top=8pt,
    bottom=4pt,
    colback=blue!6!white,
    colframe=black,
    colbacktitle=black,
    enhanced,
    center,
    attach boxed title to top left={yshift=-0.1in,xshift=0.15in},
    boxed title style={boxrule=0pt,colframe=white,},
  }
}
\newtcolorbox{AIbox}[2][]{aibox,title=#2,#1}

\begin{document}

% \title[Article Title]{HamEvo: Neural Hamiltonian Evolution Operator for Scalable and Efficient Electronic Structure Calculation}
% \title[Article Title]{HamEvo: A Neural Self-Consistent Field Operator for Transferable Kohn–Sham Hamiltonian Prediction}
% \title[Article Title]{A Fixed-Point Neural Operator for Size- and Functional-Transferable Kohn–Sham Hamiltonian Prediction}
\title[Article Title]{A Fixed-Point Neural Operator for Size- and Functional-Transferable Hamiltonian Prediction}

\author[1,2]{\fnm{Yunhong} \sur{Lou}}
\author[1]{\fnm{Xihang} \sur{Yue}}
\author[2,3]{\fnm{Xinran} \sur{Wei}}
\author*[1]{\fnm{Tianqi} \sur{Deng}}\email{dengtq@zju.edu.cn}
\author*[1]{\fnm{Linchao} \sur{Zhu}}\email{zhulinchao@zju.edu.cn}
\affil[1]{\orgname{Zhejiang University}, \orgaddress{\city{Hangzhou}, \country{China}}}
\affil[2]{\orgname{Zhongguancun Academy}, \orgaddress{\city{Beijing}, \country{China}}}
\affil[3]{\orgname{Zhongguancun Institute of Artificial Intelligence}, \orgaddress{\city{Beijing}, \country{China}}}

\abstract{
Predicting the Kohn-Sham Hamiltonian with machine learning can accelerate density functional theory while retaining access to molecular orbitals, energy levels, and electronic-structure observables that energy-only surrogates cannot resolve. Yet element-wise agreement with the converged Hamiltonian, an implicit fixed point of the self-consistent field iteration, does not determine the occupied subspace that governs orbital energies and densities. Here we present HamEvo, a neural operator that learns the single-step self-consistent update and returns the converged Hamiltonian as its fixed point. HamEvo is pre-trained on intermediate self-consistent trajectories and calibrated at equilibrium with density-matrix supervision. Across benchmarks from MD17 to drug-like QMugs, HamEvo lowers Hamiltonian errors by 35--49\% over direct-regression and deep-equilibrium baselines, and predicts QMugs HOMO and LUMO energies with mean absolute errors of 0.036 and 0.053~eV, near the 1~kcal/mol chemical-accuracy scale. Few-shot fine-tuning with only 20 reference conformations extends HamEvo to molecules of up to 122 atoms, well beyond the size range covered by pre-training. With thermal molecular-dynamics sampling, HamEvo captures temperature-dependent HOMO--LUMO gap renormalization beyond the harmonic approximation. Inference is up to $242\times$ faster than conventional DFT.
}

\maketitle

\section{Introduction}

Electronic structure calculations underpin modern computational chemistry, physics, and materials science, enabling first-principles access to orbital energies, molecular orbitals, charge densities, and related spectroscopic observables~\cite{szabo2012modern}.
Kohn-Sham density functional theory (KS-DFT)~\cite{hohenberg1964inhomogeneous, kohn1965self} is the most widely used first-principles framework for these calculations~\cite{butler2018machine, von2020exploring}, but its $O(N^3)$ computational scaling with system size creates a severe bottleneck for large molecules and high-throughput screening applications~\cite{ratcliff2017challenges}.
Considerable effort has been devoted to reducing this cost through approximate methods such as semi-empirical tight-binding schemes~\cite{xtb2021methods, GFN2-xTB2019Bannwarth}, fragment-based methods~\cite{gordon2012fragmentation}, and linear-scaling DFT approaches~\cite{Bowler_2012}.
However, achieving both the efficiency of approximate methods and the accuracy of full KS-DFT remains an open challenge.

Recently, machine learning interatomic potentials have achieved impressive accuracy in predicting energies and forces with significantly reduced computational cost~\cite{schutt2018schnet, batzner2022nequip, batatia2022mace, unke2021spooky}. However, energies and forces alone do not provide access to the underlying electronic structure, such as molecular orbitals, band gaps, or density of states, which are essential for applications in 
electronic spectroscopy, charge transport, and optoelectronic design
~\cite{butler2018machine, von2020exploring}. To address this limitation, other approaches target intermediate quantum-mechanical quantities such as electron densities~\cite{Grisafi2019, Jorgensen2022, neuralscf} and Hamiltonian matrices~\cite{schutt2019schnorb, unke2021phisnet, li2022deep, yu2023efficient}. The Hamiltonian matrix, the matrix representation of the Kohn-Sham operator encoding kinetic, nuclear-attraction, and electron-electron interaction terms, is a particularly attractive prediction target because diagonalising it directly yields orbital energies, wavefunctions, and the electron density.

Existing machine learning methods for Hamiltonian prediction typically treat the task as either a single-step mapping~\cite{schutt2019schnorb, unke2021phisnet, li2022deep, yu2023efficient, luo2025efficient, li2025enhancing} or an iterative refinement problem~\cite{wang2024infusing, kim2025highorder}, but in both cases the main supervision target is still the converged Hamiltonian. In KS-DFT, the converged Hamiltonian is not an explicit function of molecular geometry that can be obtained without self-consistent solution. It is an implicitly defined fixed point of the self-consistent field (SCF) procedure, in which the Hamiltonian and the electron density are updated together until consistency is reached. Learning a direct mapping from geometry to this fixed point therefore requires the model to absorb the full nonlinear electronic response in one step, which can reduce transferability across molecular size, conformation, and exchange-correlation functional. Moreover, small element-wise errors in the Hamiltonian do not necessarily imply accurate electronic structure, because many observables are governed by the occupied subspace and the corresponding density matrix. Training only on the final converged state also discards the intermediate SCF trajectory, which contains the update information that drives the system toward self-consistency. These considerations motivate a model that learns the Hamiltonian update rule induced by the SCF procedure and calibrates the electronic structure at the resulting fixed point, rather than regressing directly to the final Hamiltonian.

We introduce HamEvo, which formulates Hamiltonian prediction as solving for the fixed point of a neural operator $\mathcal{F}_\theta$ that mimics one update step of the SCF procedure. Given the current Hamiltonian $\mathbf{H}^{(t)}$ and molecular geometry $\mathcal{G} = (\mathbf{Z}, \mathbf{R})$, where $\mathbf{Z}$ and $\mathbf{R}$ denote atomic numbers and coordinates, $\mathcal{F}_\theta$ produces an updated Hamiltonian $\mathbf{H}^{(t+1)}$, and the predicted solution $\mathbf{H}^{\star}$ is the fixed point satisfying $\mathbf{H}^{\star} = \mathcal{F}_\theta(\mathbf{H}^{\star}; \mathcal{G})$ (Fig.~\ref{fig:architecture}). The central insight is to learn the update dynamics of the SCF process rather than regressing directly to the converged Hamiltonian. Because the SCF update rules are physically grounded and remain invariant across diverse molecular systems, they represent a more transferable learning target than system-specific equilibrium states. Unlike prior iterative approaches to Hamiltonian prediction~\cite{wang2024infusing, kim2025highorder} that update predictions without reference to SCF physics, HamEvo constrains the update dynamics through SCF trajectory supervision and the fixed point through density matrix calibration.
In the pre-training stage, $\mathcal{F}_\theta$ is supervised on SCF trajectories generated by DFT calculations to learn the single-step transition $\mathbf{H}^{(t)} \rightarrow \mathbf{H}^{(t+1)}$. This trajectory-level supervision is the key mechanism by which HamEvo acquires transferable knowledge of electronic structure evolution rather than memorizing system-specific equilibrium states. In the calibration stage, the fixed point $\mathbf{H}^{\star}$ is calibrated by matching its induced density matrix $\mathbf{P}(\mathbf{H}^{\star})$ to DFT references. This density matrix objective directly supervises the occupied orbital subspace that entrywise Hamiltonian losses cannot resolve.

HamEvo demonstrates strong generalization across chemical diversity, system scale, and exchange-correlation functional.
First, across chemical diversity, the model maintains superior accuracy from simple molecules (MD17~\cite{chmiela2017machine}, QM9~\cite{ramakrishnan2014quantum}) to complex drug-like structures (GDB17~\cite{ruddigkeit2012enumeration}, QMugs~\cite{isert2022qmugs}) in a zero-shot setting.
Second, across system scale, HamEvo is adapted to molecules significantly beyond the training distribution (up to 122 atoms) after fine-tuning with a small number of reference conformations, reaching low Hamiltonian and occupied-orbital energy errors.
Third, across exchange-correlation functionals, HamEvo adapts a B3LYP-pretrained base model to $\omega$B97X-D on the large-scale $\nabla^2$DFT benchmark and to PBE0, SCAN0, $\omega$B97X, and $\omega$B97M-V in single-molecule conformational ensembles.
Together, these results demonstrate that learning physically grounded SCF dynamics enables HamEvo to generalize beyond its training distribution in a way that direct-mapping approaches cannot.

\section{Results}

\subsection{Overview of the HamEvo Framework}

\begin{figure*}[t!]
\centering
\includegraphics[width=0.98\textwidth]{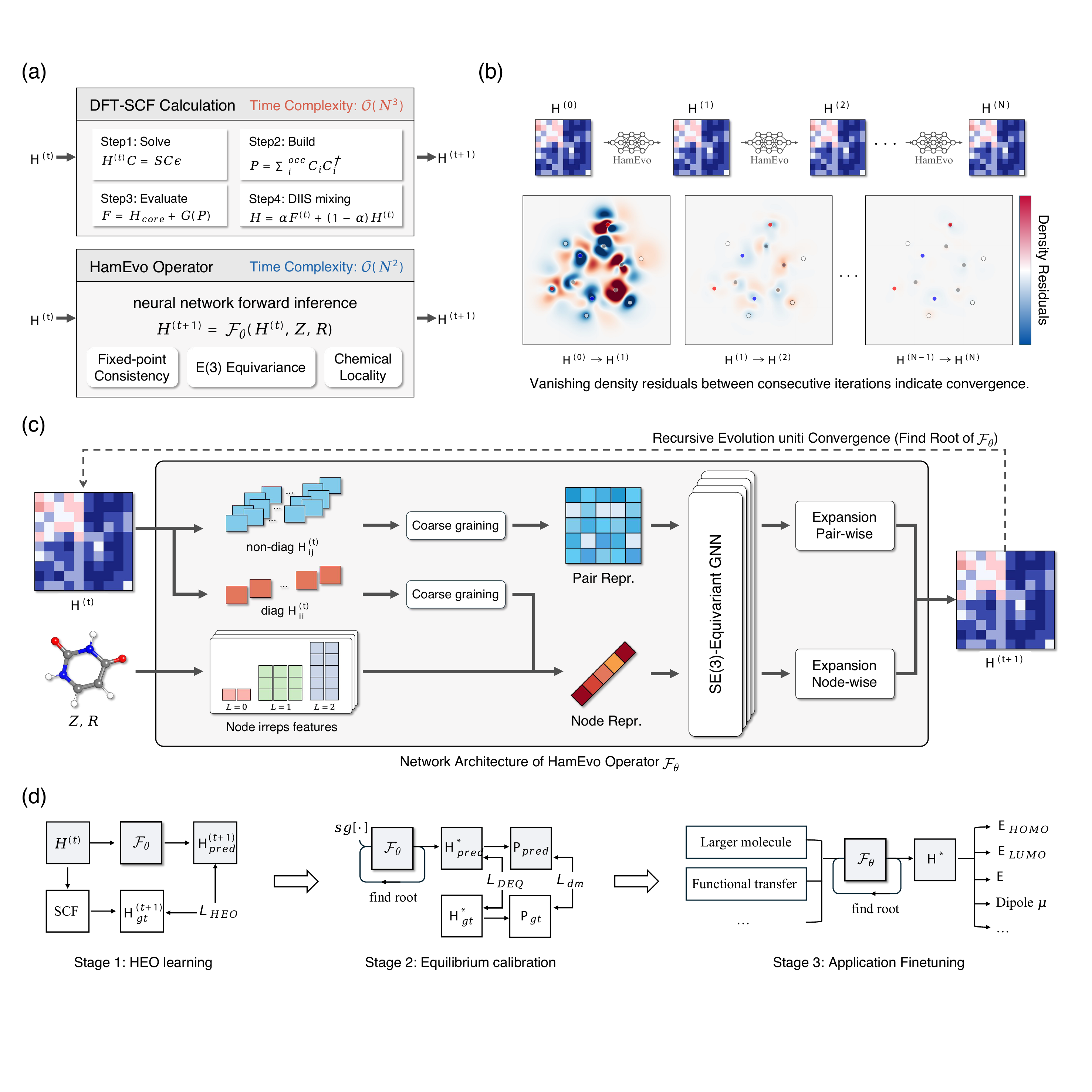}
\caption{Framework of HamEvo. \textbf{(a)} Algorithmic comparison. HamEvo reformulates the computationally expensive SCF iteration loop in DFT, which in practice scales roughly as $O(N^3)$ with system size, into an efficient ${O}(N^2)$ neural evolution operator $\mathcal{F}_{\theta}$ (bottom). The operator is designed to satisfy fixed-point consistency, E(3) equivariance, and chemical locality. \textbf{(b)} Visualization of the evolution trajectory. The model iteratively updates the Hamiltonian matrix (top) and electron density (middle) from initialization $H^{(0)}$ to convergence $H^{(N)}$. Vanishing density residuals illustrate fixed-point convergence. \textbf{(c)} Network Architecture of the HamEvo Operator $\mathcal{F}_{\theta}$. The architecture processes the current Hamiltonian $H^{(t)}$ and molecular geometry $(Z, R)$ through coarse-graining to extract node and pair representations. These are refined via an SE(3)-Equivariant GNN and expanded to predict the updated state $H^{(t+1)}$. \textbf{(d)} Multi-stage training and application workflow. The pipeline proceeds in three stages: Stage 1 learns the single-step evolution dynamics ($\mathcal{L}_{HEO}$); Stage 2 calibrates the equilibrium state using implicit differentiation ($\mathcal{L}_{EQ}$) and density matrix supervision ($\mathcal{L}_{dm}$) to ensure spectral fidelity; Stage 3 demonstrates few-shot fine-tuning for extrapolating to larger molecules and transferring across different DFT functionals to predict properties like $E_{HOMO}$ and dipole moments.}
\label{fig:architecture}
% \vskip -0.1in
\end{figure*}

HamEvo learns a neural operator that reproduces the iterative update dynamics of the SCF procedure, treating the converged Hamiltonian as the fixed point of this learned process. We first outline the framework. The core component is a learnable \textit{Hamiltonian Evolution Operator} (HEO) $\mathcal{F}_\theta$. The HEO is designed to approximate one step of the classical SCF update, mapping the current Hamiltonian estimate $\mathbf{H}^{(t)}$ and the molecular geometry $\mathcal{G}$ to an updated estimate:
\begin{equation}
\mathbf{H}^{(t+1)} = \mathcal{F}_\theta(\mathbf{H}^{(t)}; \mathcal{G}).
\end{equation}
Starting from an initial guess $\mathbf{H}^{(0)}$, the operator is applied repeatedly until the Hamiltonian reaches a fixed point $\mathbf{H}^{\star} = \mathcal{F}_\theta(\mathbf{H}^{\star}; \mathcal{G})$, which is taken as the final prediction (Fig.~\ref{fig:architecture}a).
Fig.~\ref{fig:architecture}(b) provides a qualitative view of how HamEvo approaches self-consistency during inference. We diagonalize each intermediate Hamiltonian and visualize the real-space density residual between consecutive iterations, whose attenuation illustrates the convergence behavior.

The architecture of the HEO follows four stages, including geometric graph construction, Hamiltonian-state projection, equivariant message passing, and block reconstruction (Fig.~\ref{fig:architecture}(c)). Under a fixed atomic-orbital (AO) basis, the Hamiltonian naturally decomposes into atom-pair blocks, where diagonal blocks $\mathbf{H}_{ii}$ capture on-site terms and off-diagonal blocks $\mathbf{H}_{ij}$ capture inter-atomic couplings. We first construct a directed graph over ordered atom pairs from the molecular geometry and initialize node features from atom-type embeddings plus geometry-derived edge-degree embeddings. The current Hamiltonian state $\mathbf{H}^{(t)}$ is then projected into SO(3)-equivariant latent features by separate diagonal and off-diagonal modules, which map the atom-pair blocks into a shared latent representation. These state features are injected into a stack of equivariant transformer blocks, where geometric edge features guide message passing. Finally, the refined latent states are expanded back to diagonal and off-diagonal Hamiltonian blocks, symmetrized, and masked to produce $\mathbf{H}^{(t+1)}$. The detailed network architecture is described in Sec.~\ref{method:archi}.

Training proceeds in three stages (Fig.~\ref{fig:architecture}d). Stage~1 learns the single-step SCF dynamics, Stage~2 calibrates the equilibrium fixed point with density-matrix supervision, and Stage~3 adapts the pre-trained HamEvo model to downstream applications. Detailed loss definitions and optimization procedures are provided in Sec.~\ref{method:training}, and dataset construction and DFT settings are described in Sec.~\ref{method:datasets}.

In Stage~1 (\textit{HEO Learning}), we record full SCF convergence trajectories and extract consecutive Hamiltonian pairs $(\mathbf{H}^{(t)}, \mathbf{H}^{(t+1)})$. The model is trained with the one-step evolution objective $\mathcal{L}_{\text{HEO}}$ to reproduce individual SCF update steps.

In Stage~2 (\textit{Equilibrium Calibration}), we optimize the fixed-point Hamiltonian $\mathbf{H}^{\star}$ against the DFT-converged Hamiltonian $\mathbf{H}^{\text{c}}$ and its corresponding density matrix $\mathbf{P}^{\text{c}}$. This stage combines an equilibrium Hamiltonian loss $\mathcal{L}_{\text{EQ}}$ with density-matrix supervision $\mathcal{L}_{\text{dm}}$, using implicit differentiation through the fixed point~\cite{bai2019deep}.

In Stage~3 (\textit{Application Fine-tuning}), the pre-trained HamEvo model is adapted to specific downstream tasks. Although the SCF update dynamics learned in Stages 1--2 are broadly transferable, molecules substantially larger than the training set or computed under different exchange-correlation functionals introduce electronic environments not fully covered by pre-training. Task-specific fine-tuning efficiently bridges this gap, as demonstrated in Sec.~\ref{sec:results_transfer} and Sec.~\ref{sec:results_functional}.

\subsection{Robust Predictions Across Scales}
\label{sec:results_robust_predictions}

We first evaluate HamEvo across molecular benchmarks of increasing structural complexity. The evaluation uses four datasets ranging from simple conformational variations to complex drug-like molecules, including MD17-ethanol~\cite{chmiela2017machine}, QH9~\cite{ramakrishnan2014quantum}, GDB17~\cite{ruddigkeit2012enumeration}, and QMugs~\cite{isert2022qmugs} (Fig.~\ref{fig:general_results}a; Methods~\ref{method:datasets}). For the GDB17/QMugs benchmark, the dataset partition is described in Methods.
We compare HamEvo to two baselines representing the two main prediction paradigms: (i) Direct Regression (denoted as \textbf{Reg})~\cite{schutt2019unifying}, which maps the molecular structure to the converged Hamiltonian in a single forward pass; and (ii) Deep Equilibrium Models (denoted as \textbf{DEQ})~\cite{wang2024infusing}, which enforce self-consistency through implicit differentiation.
We use Hamiltonian MAE, derived-property errors, and occupied-orbital similarity ${C_{\text{sim}}}$ as evaluation metrics. Derived properties are computed as described in Methods~\ref{method:evaluation}, with full metric definitions provided in Appendix~\ref{appendix:metric_define}.

HamEvo achieves lower Hamiltonian MAEs across all four benchmarks.
As summarized in Fig.~\ref{fig:general_results}b, HamEvo outperforms the best baseline on every dataset, reducing the MAE by 35--49\% (Table~\ref{tab:hamiltonian_mae}). Even for the complex drug-like molecules in the QMugs test benchmark, the MAE remains at 1.30~meV, which is comparable to the errors on smaller benchmarks and indicates that prediction accuracy is maintained within the evaluated size range.
These consistent improvements across all four benchmarks support the effectiveness of modeling iterative Hamiltonian evolution for capturing the nonlinear geometry-to-Hamiltonian relationship, compared to single-step regression and equilibrium-only approaches.

Beyond numerical accuracy of matrix elements, HamEvo yields physically relevant electronic properties.
We evaluated derived properties on the QMugs test set of 9,277 conformations, including $E_{\mathrm{HOMO}}$, $E_{\mathrm{LUMO}}$, dipole moment $\mu$, and occupied-orbital similarity ${C_{\text{sim}}}$ (Fig.~\ref{fig:general_results}c; Table~\ref{tab:derived_property_accuracy}; Methods~\ref{method:datasets}).
HamEvo predicts $E_{\mathrm{HOMO}}$ and $E_{\mathrm{LUMO}}$ with MAEs of 0.036~eV and 0.053~eV, respectively. For reference, these values are on the order of the widely cited 1~kcal/mol threshold, approximately 0.043~eV. Additionally, HamEvo achieves a dipole moment MAE of 0.098~Debye.
HamEvo maintains a mean orbital similarity ${C_{\text{sim}}}$ of 0.974 for occupied orbital coefficients.
Together, these results show that HamEvo's matrix-level accuracy translates into reliable derived electronic observables.

To complement these aggregate benchmarks, we analyzed Morphine ($\mathrm{C}_{17}\mathrm{H}_{19}\mathrm{N}\mathrm{O}_{3}$), a held-out pharmaceutical compound, as a representative case study (Fig.~\ref{fig:general_results}d).
To localize the matrix errors, we grouped atom-pair Hamiltonian blocks by the corresponding interatomic distance $R_{ij}$ and computed the MAE within each distance bin. HamEvo shows the largest error reduction in the short-range region associated with covalent bonding ($R < 2.2$~\AA), including a 73.9\% lower block MAE relative to the best baseline (Fig.~\ref{fig:general_results}f).
The predicted HOMO and LUMO isosurfaces closely match the DFT references, with ${C_{\text{sim}}}$ values of 0.98 and 0.97 (Fig.~\ref{fig:general_results}e).
At the spectrum level, the kernel-density estimate (KDE) of orbital energies and discrete orbital sticks show that HamEvo follows the DFT distribution and preserves the occupied-virtual separation. It reproduces the HOMO--LUMO gap within 0.08~eV (5.23~eV vs.\ 5.31~eV), whereas the baselines introduce spurious in-gap states (Fig.~\ref{fig:general_results}g).

Together, these benchmark and case-study results show that HamEvo maintains accurate Hamiltonian predictions across molecular scales while preserving the derived electronic structure. This supports the iterative evolution formulation as a robust approach for molecular Hamiltonian prediction.

\subsection{Data-efficient Transfer to Larger Molecules}
\label{sec:results_transfer}

\begin{figure*}[t!]
\centering
\includegraphics[width=0.98\textwidth]{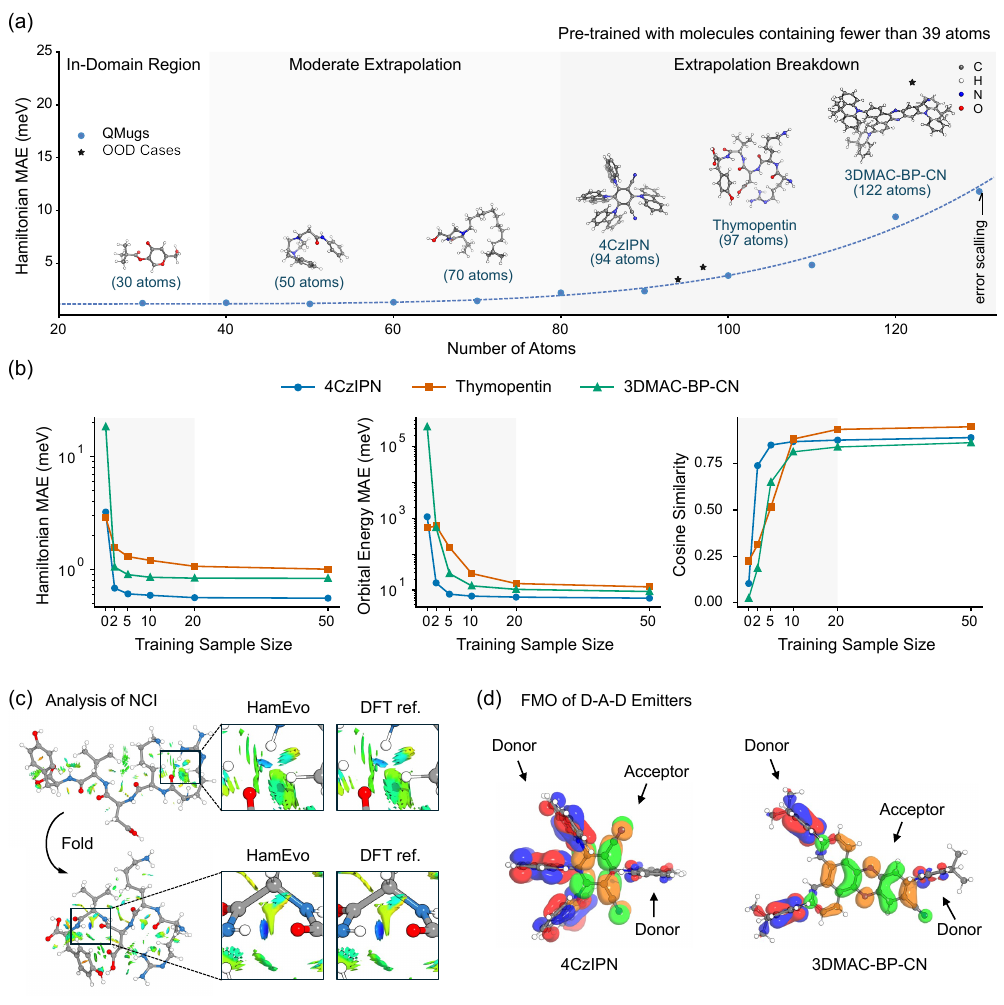}
% \vskip -0.1in
\caption{Data-efficient transfer to large-scale molecular systems. \textbf{(a)} Zero-shot generalization error versus system size. The model was pre-trained on the small-molecule training split described in Methods. The dotted line shows the empirical size-dependent error trend, with larger errors beyond approximately 80 atoms. Three external OOD targets are marked separately from the QMugs-derived trend: 4CzIPN (94 atoms, rigid emitter), Thymopentin (97 atoms, flexible peptide), and 3DMAC-BP-CN (122 atoms). \textbf{(b)} Few-shot fine-tuning performance on the three representative molecules. Left to right: Hamiltonian MAE, mean absolute error of all occupied orbital energies, and cosine similarity between predicted and DFT reference orbitals. The shaded region marks fine-tuning with up to 20 reference conformations. The dashed line marks the room-temperature thermal energy scale ($k_\mathrm{B}T \approx 26$~meV). \textbf{(c)} Non-covalent interaction (NCI) analysis of Thymopentin after fine-tuning with 20 reference conformations. HamEvo and DFT reference visualizations are shown for two folded conformations. Zoomed insets focus on hydrogen-bonding regions, where blue NCI isosurfaces indicate hydrogen-bond interactions. \textbf{(d)} Frontier molecular orbitals (FMO) of donor-acceptor-donor (D-A-D) emitters after the same 20-shot fine-tuning. Red/blue isosurfaces denote the HOMO, and orange/green isosurfaces denote the LUMO, showing spatial separation across donor and acceptor regions in 4CzIPN and 3DMAC-BP-CN.}
\label{fig:finetune_results}
% \vskip -0.2in
\end{figure*}

We first evaluated the pre-trained HamEvo model on molecules of increasing size in a zero-shot setting (Fig.~\ref{fig:finetune_results}a).
This test identifies the onset of size-dependent error growth and motivates few-shot adaptation to larger target molecules.
As shown in Fig.~\ref{fig:finetune_results}a, the zero-shot error is smallest within the size range covered by the pre-training data and increases as molecular size extends beyond it. Beyond 80 atoms, the error grows more sharply, marking the size range used for the few-shot experiments.
We then focused on three external out-of-distribution (OOD) targets marked separately from the QMugs-derived size trend in Fig.~\ref{fig:finetune_results}a: 4CzIPN ($\mathrm{C}_{56}\mathrm{H}_{32}\mathrm{N}_{6}$, 94 atoms), Thymopentin ($\mathrm{C}_{30}\mathrm{H}_{49}\mathrm{N}_{9}\mathrm{O}_{9}$, 97 atoms), and 3DMAC-BP-CN ($\mathrm{C}_{66}\mathrm{H}_{50}\mathrm{N}_{6}$, 122 atoms). They represent a rigid conjugated fluorescent emitter, a flexible peptide drug, and a strongly conjugated organic system, respectively.
For each molecule, we generated a conformational ensemble by uniformly sampling 2,000 structures from 100~ps molecular dynamics simulations at 300~K using xTB-GFN0.
We computed reference B3LYP/def2-SVP Hamiltonians for these structures to evaluate adaptation with a limited number of reference conformations.

Using these target ensembles, we fine-tuned HamEvo with increasing numbers of reference conformations and evaluated the resulting prediction accuracy (Fig.~\ref{fig:finetune_results}b).
The zero-shot setting gives Hamiltonian MAEs of 3.24, 2.90, and 18.32~meV for 4CzIPN, Thymopentin, and 3DMAC-BP-CN, respectively.
Fine-tuning reduces these errors mainly within the first 20 reference conformations, reaching 0.57, 1.07, and 0.84~meV at 20 shots.
Increasing the reference set to 50 conformations gives similar MAEs of 0.56, 1.01, and 0.84~meV, suggesting that most matrix-level improvement is already obtained by 20 shots.

Beyond matrix-level accuracy, we also evaluated occupied-orbital energy errors and orbital coefficient agreement after fine-tuning (Fig.~\ref{fig:finetune_results}b).
The orbital-level metrics show the same data-efficient trend.
At 20 reference conformations, occupied-orbital energy MAEs are 6.34--15.10~meV across the three targets, below the room-temperature thermal energy scale ($k_\mathrm{B}T \approx 26$~meV).
The corresponding orbital coefficient cosine similarities are 0.84--0.94 at 20 shots and 0.86--0.95 at 50 shots.

As illustrative case studies, we examine electronic-structure features in the three target molecules after 20-shot fine-tuning. For Thymopentin, Fig.~\ref{fig:finetune_results}c compares HamEvo and DFT reference NCI visualizations for two folded conformations. The zoomed regions show hydrogen-bonding interactions with similar NCI isosurface patterns.
For the conjugated emitters 4CzIPN and 3DMAC-BP-CN, Fig.~\ref{fig:finetune_results}d shows the predicted frontier molecular orbitals. The HOMO and LUMO isosurfaces are spatially separated across donor and acceptor regions, consistent with the expected D-A-D electronic structure.

In summary, zero-shot errors are smallest near the size range covered by the pre-training data and increase beyond approximately 80 atoms. Fine-tuning with 20 reference conformations reduces Hamiltonian errors for three target molecules up to 122 atoms. This 20-shot adaptation also reduces occupied-orbital energy errors and increases orbital coefficient similarity. The case studies provide qualitative checks on NCI density patterns and D-A-D frontier-orbital separation after adaptation.

\subsection{Cross-Functional Adaptation}
\label{sec:results_functional}

\begin{figure*}[t!]
\centering
\includegraphics[width=0.98\textwidth]{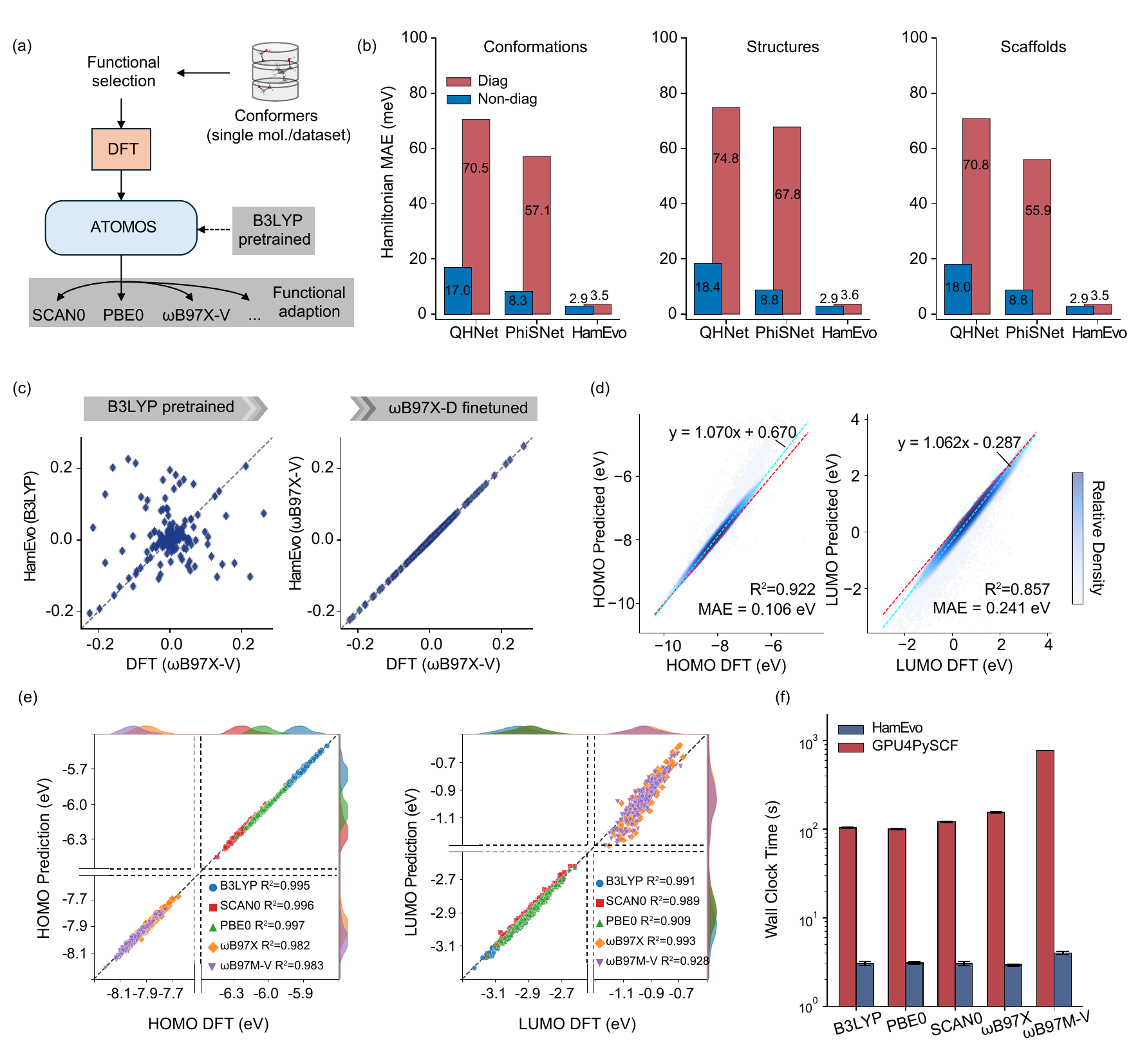}
% \vskip -0.1in
\caption{Cross-functional adaptation and computational efficiency.
\textbf{(a)} Schematic of the transfer learning workflow. HamEvo pre-trained on B3LYP data is adapted to target functionals via fine-tuning.
\textbf{(b)} Dataset-level transfer on the $\nabla^2$DFT benchmark ($\omega$B97X-D). Hamiltonian MAEs are reported separately for diagonal and off-diagonal blocks across Conformation, Structure, and Scaffold splits.
\textbf{(c)} Hamiltonian-entry scatter before and after fine-tuning from B3LYP to $\omega$B97X-D. The x-axis gives target-functional DFT entries, and the y-axis gives HamEvo predictions.
\textbf{(d)} HOMO and LUMO energy prediction on the Structure split (400k conformations), reported with $R^2$ and MAE.
\textbf{(e)} Few-shot adaptation for 4CzIPN across five functionals. Left and right panels show predicted versus DFT HOMO and LUMO energies, respectively; colors denote functionals, and marginal densities show the corresponding energy distributions.
\textbf{(f)} Wall-clock time per conformation on a logarithmic scale. HamEvo runtimes are compared with the corresponding GPU4PySCF DFT runtimes.}
\label{fig:functional_results}
\vskip -0.2in
\end{figure*}

In practical DFT studies, the choice of exchange-correlation functional changes the target Hamiltonian and the derived orbital energies. Retraining a separate Hamiltonian model for each functional would therefore be costly.
We evaluate whether a B3LYP-pretrained HamEvo model can be adapted to other functionals in two complementary settings (Fig.~\ref{fig:functional_results}a).
First, we test dataset-level transfer on the large-scale $\nabla^2$DFT benchmark, where the model is fine-tuned to the target $\omega$B97X-D functional and evaluated across Conformation, Structure, and Scaffold splits (Fig.~\ref{fig:functional_results}b--d).
Second, we test single-molecule multi-functional adaptation on 4CzIPN using PBE0, SCAN0, $\omega$B97X, and $\omega$B97M-V as target functionals. These functionals span global hybrid, hybrid meta-GGA, and range-separated hybrid approximations (Fig.~\ref{fig:functional_results}e,f).
The dataset-level benchmark tests whether functional adaptation scales across chemical space, whereas the single-molecule setting isolates functional-specific shifts under a fixed conformational ensemble.
For the dataset-level $\nabla^2$DFT benchmark, we compare HamEvo with QHNet~\cite{yu2023efficient} and PhiSNet~\cite{unke2021se} baselines trained from scratch on the target-functional data.

On the dataset-level $\nabla^2$DFT benchmark, Fig.~\ref{fig:functional_results}b compares diagonal and off-diagonal Hamiltonian MAEs across the Conformation, Structure, and Scaffold splits described in Methods~\ref{method:datasets}.
This separation is informative because off-diagonal blocks contain many weak interatomic couplings, whereas diagonal blocks encode on-site AO terms with larger element- and environment-dependent variations.
HamEvo gives diagonal MAEs of 3.5--3.6~meV and off-diagonal MAEs of 2.9~meV across the three splits.
The corresponding best baseline errors are 55.9--67.8~meV for diagonal blocks and 8.3--8.8~meV for off-diagonal blocks.
The largest difference appears in the diagonal Hamiltonian blocks, indicating that the improvement is not limited to the weak off-diagonal couplings but also extends to on-site AO terms.

HamEvo also addresses the systematic offset between functionals at the Hamiltonian-entry level.
As visualized in Fig.~\ref{fig:functional_results}c, directly applying the B3LYP-pretrained model to $\omega$B97X-D data results in a large systematic deviation, with predicted matrix entries scattered off the diagonal (left panel). Fine-tuning reduces this distribution shift, bringing the predictions closer to the identity line (right panel).
This matrix-level correction carries over to frontier orbital energies: on the Structure split (400k conformations), HamEvo predicts HOMO and LUMO energies with $R^2 > 0.85$ and MAEs of 0.106~eV and 0.241~eV, respectively (Fig.~\ref{fig:functional_results}d). While the LUMO errors are larger, reflecting the well-known difficulty of predicting unoccupied states, these correlations support the utility of this approach for frontier-energy screening at this benchmark scale.

To probe whether the model can distinguish between functionals for a single molecular system, we fine-tuned HamEvo on the conjugated molecule 4CzIPN with 20 reference conformations from each of four additional functionals (SCAN0, PBE0, $\omega$B97X, $\omega$B97M-V).
Different functionals shift frontier energies by up to 2.4~eV, and the fine-tuned model distinguishes these functional-dependent energy shifts (Fig.~\ref{fig:functional_results}e). All HOMO predictions achieve $R^2 > 0.98$, while LUMO correlations remain high ($R^2 \approx 0.99$ for global hybrids and $>0.91$ for range-separated functionals).

HamEvo also decouples inference cost from functional complexity.
Fig.~\ref{fig:functional_results}f compares the wall-clock time per conformation on a logarithmic scale.
While conventional DFT calculations vary drastically with functional complexity, ranging from 100.6~s (PBE0) to 775.4~s ($\omega$B97M-V) per conformation, HamEvo shows only minor variation across functionals, with mean inference times ranging from 2.927 to 3.998~s.
All computational timings were measured on a single NVIDIA RTX 4090 GPU.
This represents a speedup of up to 242$\times$ compared to standard DFT, making high-throughput screening with advanced functionals computationally feasible.

Together, these results show that a single pre-trained Hamiltonian model can be adapted across exchange-correlation functionals in both large-scale benchmark transfer and few-shot single-molecule settings. The adapted model retains functional-specific electronic signatures while reducing the cost of obtaining Hamiltonian predictions at alternative levels of theory.

\subsection{Band Gap Renormalization}

Beyond static equilibrium properties, accurate modeling of molecular systems under finite-temperature conditions requires capturing the impact of thermal fluctuations on electronic structure. Here we use gap renormalization to denote the temperature-dependent shift of the HOMO--LUMO gap arising from nuclear quantum and thermal fluctuations, by analogy with the band gap renormalization (BGR) framework established for periodic systems. This effect, arising from electron-phonon coupling and further complicated by anharmonicity, necessitates extensive statistical sampling that is often prohibitively expensive for standard DFT~\cite{hohenberg1964inhomogeneous, kohn1965self, ratcliff2017challenges}. We demonstrate HamEvo's capability by performing finite-temperature simulations on two distinct systems: Pentamantane ($\mathrm{C}_{26}\mathrm{H}_{32}$), a rigid wide-gap insulator, and NAI-DMAC ($\mathrm{C}_{31}\mathrm{H}_{24}\mathrm{N}_{2}\mathrm{O}$), a flexible thermally activated delayed fluorescence (TADF) emitter.

\begin{figure*}[t!]
\centering
\includegraphics[width=0.98\textwidth]{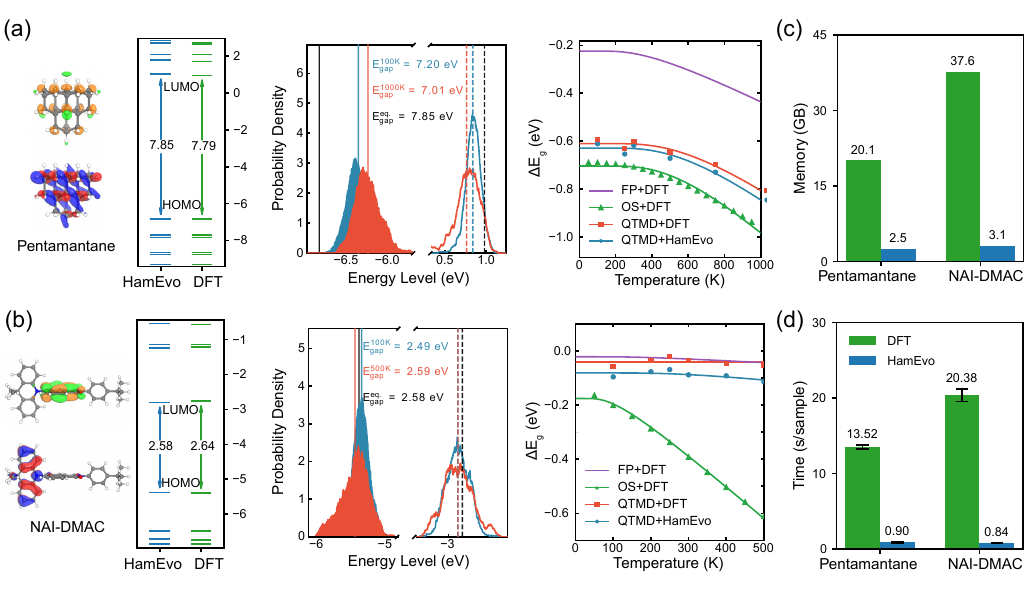}
% \vskip -0.1in
\caption{Temperature-dependent HOMO--LUMO gap renormalization.
\textbf{(a)} Pentamantane results. Left to right: frontier molecular orbitals and equilibrium HOMO--LUMO levels from HamEvo (blue) and DFT (green), probability densities of HOMO and LUMO energies sampled at 100~K and 1000~K, and the temperature-dependent gap shift $\Delta E_{\mathrm{gap}}(T)$. In the orbital-energy distributions, vertical guide lines mark the frontier-orbital levels used to compute the displayed gaps.
\textbf{(b)} NAI-DMAC results in the same format, with orbital-energy distributions shown at 100~K and 500~K.
In the $\Delta E_{\mathrm{gap}}(T)$ plots, quantum thermal molecular dynamics with HamEvo (QTMD+HamEvo) is compared with QTMD+DFT, frozen-phonon DFT (FP+DFT), and one-shot displacement DFT (OS+DFT).
\textbf{(c)} Average memory consumption for DFT and HamEvo.
\textbf{(d)} Wall-clock time per sample for DFT and HamEvo.}
\label{fig:EP_renormalization}
\vskip -0.2in
\end{figure*}

First, we verify the model's accuracy on static equilibrium structures. As shown in Fig.~\ref{fig:EP_renormalization}a,b, HamEvo accurately reproduces the frontier molecular orbitals and energy levels for these two systems. For Pentamantane, the predicted HOMO--LUMO gap is 7.85~eV, virtually identical to the DFT reference of 7.79~eV. For the conjugated NAI-DMAC system, HamEvo predicts a gap of 2.58~eV (DFT: 2.64~eV) and correctly reconstructs the charge-transfer character of the orbitals, with the HOMO localized on the donor (DMAC) and the LUMO on the acceptor (NAI).

A key advantage of HamEvo in this context is its computational efficiency for high-throughput sampling. As illustrated in Fig.~\ref{fig:EP_renormalization}d, HamEvo reduces the inference time per sample to 0.90~s for Pentamantane and 0.84~s for NAI-DMAC, representing a speedup of over $15\times$ compared to standard DFT calculations using GPU4PySCF~\cite{sun2020recent} (13.52~s and 20.38~s, respectively). This efficiency facilitates the rapid accumulation of statistics required for converging thermodynamic properties.

We leveraged this speed to compute the temperature-dependent electronic structure via molecular dynamics. The middle panels of Fig.~\ref{fig:EP_renormalization}a,b display the probability densities of frontier orbital energies sampled at elevated temperatures (1000~K for Pentamantane, 500~K for NAI-DMAC). The HamEvo distributions closely match the ground-truth quantum thermal molecular dynamics (QTMD) references, capturing both the thermal broadening and the systematic shift of the energy levels relative to the equilibrium state ($E_{\mathrm{gap}}^{\mathrm{eq}}$). For Pentamantane, the thermally averaged gap shrinks to $E_{\mathrm{gap}}^{1000\,\mathrm{K}} = 7.20$~eV from the equilibrium 7.85~eV, a significant renormalization effect that HamEvo reproduces accurately.

Finally, we quantify the gap renormalization $\Delta E_{\mathrm{gap}}(T)$ across a wide temperature range in the right panels of Fig.~\ref{fig:EP_renormalization}a,b. HamEvo (blue circles) exhibits excellent agreement with the QTMD benchmarks (red squares), accurately describing the monotonic decrease of the HOMO--LUMO gap with increasing temperature. Crucially, HamEvo outperforms the harmonic Frozen Phonon (FP, purple line) and One-Shot special displacement (OS, green line) approximations, which fail to capture anharmonic effects at higher temperatures. These results demonstrate that HamEvo, combined with molecular dynamics sampling, captures anharmonic gap renormalization effects that harmonic approximations miss and therefore provides a proof of concept for accelerating finite-temperature electronic-structure calculations in molecular systems.

\section{Discussion}

In this study, we introduce HamEvo, a framework that formulates Hamiltonian prediction as \emph{iterative evolution} rather than \emph{direct regression}~\cite{schutt2019unifying,unke2021phisnet,li2022deep,yu2023efficient}. Direct regression attempts to learn a static mapping from geometry to the Hamiltonian, but this mapping is inherently sensitive to global structural variations~\cite{ratcliff2017challenges,kummel2008orbital}. HamEvo instead learns a neural evolution operator that mimics the SCF procedure~\cite{hohenberg1964inhomogeneous,kohn1965self}. This formulation helps the model capture the invariant physical rules of electron interactions (such as Coulomb and exchange forces) rather than merely fitting the specific numerical values of equilibrium states. Consequently, HamEvo achieves robust transferability, successfully extrapolating to systems three times larger than those in the training set and adapting to diverse functionals with minimal data.

HamEvo shares with NeuralSCF the use of intermediate SCF trajectories to supervise single-step updates and implicit differentiation for fixed-point calibration~\cite{neuralscf}. It differs in treating the AO Hamiltonian rather than the electron density as the evolving state. This choice makes orbital-resolved quantities native to the predicted operator. Solving the generalized eigenvalue problem directly yields the full orbital spectrum and molecular wavefunctions without reconstructing the Kohn--Sham operator from a predicted electron density. The atom-pair block structure also provides a natural representation of on-site terms and interatomic couplings for equivariant modeling.

The observed generalization across chemical space of HamEvo comes from decomposing the difficult, nonlinear mapping into a sequence of small, easier-to-learn steps. Direct prediction models~\cite{li2022deep,gong2023general,yu2023efficient} often struggle with large systems because a small change in structure can cause complex, long-range changes in the electronic cloud (polarization), making the global function too hard to approximate~\cite{unke2021spooky,merchant2023scaling}. In contrast, HamEvo does not predict the final answer all at once. Instead, it predicts the local correction ($\Delta \mathbf{H}$) needed to update the current state. These step-wise updates are much more local and have smaller numerical variations than the absolute Hamiltonian matrix, making them significantly easier to learn. Furthermore, training the model along SCF-like evolution paths encourages intermediate states to remain aligned with self-consistent updates~\cite{wang2024infusing}. In our large-molecule experiments, the adapted iterative process converges stably for unseen molecules up to 122 atoms.

Beyond numerical accuracy, the integration of density matrix supervision is the key reason HamEvo can correctly predict experimental properties. Conventional deep learning approaches~\cite{yu2023efficient,wang2024deeph} typically optimize the element-wise mean squared error (MSE) of the Hamiltonian matrix. However, achieving low numerical error does not automatically guarantee accurate physical observables, as properties like orbital energies are highly sensitive to the eigenvectors, particularly near the Fermi level~\cite{li2024neural}. Small numerical deviations in these critical regions can lead to incorrect orbital ordering and electron filling. By explicitly including the density matrix in the loss function, consistent with recent physics-informed strategies~\cite{hu2024self,zhang2024self,li2024neural}, HamEvo aligns the predicted eigenspace with the ground truth. This forces the model to capture the correct electron density distribution rather than merely fitting matrix entries, thereby preserving crucial spectral properties like the HOMO--LUMO gap.

While HamEvo represents a significant step forward, it has limitations common to machine learning methods. First, although the iterative inference is significantly faster than standard DFT (up to $242\times$ speedup), it is naturally slower than single-step regression models~\cite{unke2021phisnet,li2022deep}. This is a necessary cost to achieve physical robustness and generalization. Second, the current implementation uses a fixed AO basis set (def2-SVP~\cite{weigend2005balanced}). Extending it to handle different basis sets remains a challenge. Additionally, while the model can correct the physical differences between functionals (e.g., B3LYP~\cite{becke1993density} to $\omega$B97X-D~\cite{chai2008long}), its ultimate accuracy depends on the quality of the DFT training data~\cite{khrabrov2022nabladft}. Future work will focus on integrating force prediction to enable molecular dynamics simulations~\cite{batzner2022nequip,kovacs2023mace} and exploring ways to reduce the reliance on expensive high-level reference data~\cite{mathiasen2024reducing}.

In conclusion, HamEvo demonstrates that learning the process of reaching a solution is more scalable than learning the final solution directly. The framework's ability to work across different chemical spaces, system sizes, and theoretical frameworks suggests that it has learned transferable physical interactions rather than just statistical patterns. By providing DFT-level accuracy with lower inference cost, HamEvo opens the door to high-throughput screening of massive molecular systems~\cite{isert2022qmugs} and long-timescale simulations that were previously impossible.

\section{Method}

\subsection{Hamiltonian Evolution Operator}
\label{method:HEO}

In the linear combination of atomic orbitals (LCAO) ansatz, the KS-DFT problem is solved via the Roothaan-Hall equations \cite{roothaan1951new, szabo2012modern}, which creates a generalized eigenvalue problem:
\begin{equation} \label{eq:roothaan}
\mathbf{H}\mathbf{C} = \mathbf{S}\mathbf{C}\boldsymbol{\varepsilon},
\end{equation}
where $\mathbf{H}$ is the Kohn-Sham Hamiltonian (Fock) matrix, $\mathbf{S}$ is the atomic-orbital overlap matrix, $\mathbf{C}$ collects the molecular-orbital expansion coefficients, and $\boldsymbol{\varepsilon}$ contains the orbital energies.
The Hamiltonian matrix $\mathbf{H}$ depends nonlinearly on the density matrix $\mathbf{P}$, which is constructed from the occupied eigenvectors in $\mathbf{C}$. Because of this dependence, Eq. \eqref{eq:roothaan} must be solved iteratively. For a molecular geometry $\mathcal{G} = (\mathbf{Z}, \mathbf{R})$, we denote one classical SCF update by the operator $\mathcal{T}_{\text{SCF}}$:
\begin{equation} \label{eq:scf_operator}
\mathbf{H}^{(t+1)} = \mathcal{T}_{\text{SCF}}(\mathbf{H}^{(t)}; \mathcal{G}).
\end{equation}
The operator $\mathcal{T}_{\text{SCF}}$ represents one SCF update, including Hamiltonian diagonalization, density-matrix construction, and Fock matrix evaluation~\cite{szabo2012modern}.

\textbf{Formulation.}
HamEvo replaces the analytic SCF update with a learned operator $\mathcal{F}_\theta$ that maps the current Hamiltonian state to the next one:
\begin{equation} \label{eq:heo_update}
\mathbf{H}^{(t+1)} = \mathcal{F}_\theta(\mathbf{H}^{(t)}; \mathcal{G}).
\end{equation}
The model is trained as a state-to-state map rather than as a direct geometry-to-Hamiltonian regressor. In the implementation, $\mathbf{H}^{(t)}$ is represented by atom-centered diagonal blocks $\{\mathbf{H}_{ii}^{(t)}\}$ and ordered off-diagonal blocks $\{\mathbf{H}_{ij}^{(t)}\}_{i\neq j}$. The block representation is only an implementation choice, and $\mathcal{F}_\theta$ still updates the full Hamiltonian state. The converged prediction is the fixed point of the learned update:
\begin{equation} \label{eq:fixed_point}
\mathbf{H}^{\star} = \mathcal{F}_\theta(\mathbf{H}^{\star}; \mathcal{G}).
\end{equation}
At inference time, we obtain this equilibrium by iterating $\mathcal{F}_\theta$ with a fixed-point solver, rather than by predicting $\mathbf{H}^{\star}$ in one forward pass.

\textbf{Physical Constraints.}
The learned operator must satisfy three structural requirements. First, it must admit a stable fixed point, because both training and inference optimize or solve for $\mathbf{H}^{\star}$. Second, the update should respect geometric symmetry. We use a standard SO(3)-equivariant design based on relative coordinates and spherical harmonics. Third, it must preserve the algebraic structure of the Hamiltonian. In practice, we enforce this by predicting diagonal and off-diagonal atom-pair blocks separately, symmetrizing the outputs, and masking basis entries that are not valid for a given atom type.

\textbf{Discussion.}
The main modeling choice is to learn an SCF-like update map instead of regressing the converged Hamiltonian directly. This choice does not remove the need to solve for self-consistency, but it makes the network responsible for one physically structured refinement step. The fixed-point solver then composes these learned updates into the final equilibrium prediction.

\subsection{HamEvo Architecture}
\label{method:archi}

HamEvo implements $\mathcal{F}_\theta$ through four stages: geometric graph construction, Hamiltonian-state projection, equivariant message passing, and block reconstruction.

\textbf{Geometry graph and initial node features.}
For a molecule with atoms $\mathbf{Z}=\{Z_i\}_{i=1}^M$ and positions $\mathbf{R}=\{\mathbf{r}_i\}_{i=1}^M$, we construct a directed molecular graph using a distance cutoff. Each edge $(j,i)$ carries two geometric descriptors: a radial distance embedding $\boldsymbol{\rho}(r_{ij})$ and spherical harmonics $\mathbf{Y}(\hat{\mathbf{r}}_{ij})$ up to $\ell=4$. The initial node state is obtained from an atom-type embedding plus an edge-degree embedding aggregated from neighboring geometry:
\begin{equation}
\mathbf{x}_i^{(0)} = \mathrm{Emb}(Z_i) + \Psi\!\left(\left\{\boldsymbol{\rho}(r_{ij}), \mathbf{Y}(\hat{\mathbf{r}}_{ij})\right\}_{j\neq i}\right).
\end{equation}
This step depends only on geometry and atomic identity. The current Hamiltonian state is injected separately in the next stage.

\textbf{Hamiltonian-state projection.}
Under a fixed AO basis, the Hamiltonian naturally decomposes into atom-pair blocks:
\begin{equation}
\mathbf{H} = \begin{pmatrix} 
\mathbf{H}_{11} & \mathbf{H}_{12} & \cdots & \mathbf{H}_{1M} \\
\mathbf{H}_{21} & \mathbf{H}_{22} & \cdots & \mathbf{H}_{2M} \\
\vdots & \vdots & \ddots & \vdots \\
\mathbf{H}_{M1} & \mathbf{H}_{M2} & \cdots & \mathbf{H}_{MM} 
\end{pmatrix}     
\end{equation}
where $\mathbf{H}_{ij} \in \mathbb{R}^{n_i \times n_j}$ denotes the interaction between basis functions centered on atoms $i$ and $j$. We handle element-dependent block sizes using padding and binary masks. The current Hamiltonian is projected into equivariant latent features by separate diagonal and off-diagonal modules:
\begin{equation}
\mathbf{u}_{ii} = \Pi_{\mathrm{diag}}(\mathbf{H}_{ii}; \boldsymbol{\phi}_i),
\end{equation}
\begin{equation}
\mathbf{u}_{ij} = \Pi_{\mathrm{off}}(\mathbf{H}_{ij}; \boldsymbol{\phi}_{ij}),
\end{equation}
where $\boldsymbol{\phi}_i=\mathbf{x}_i^{(0)}$ is the atom-level conditioning feature, and $\boldsymbol{\phi}_{ij}=[\mathbf{x}_i^{(0)} \Vert \mathbf{x}_j^{(0)}]$ is the pair-level conditioning feature. Concretely, $\Pi_{\mathrm{diag}}$ and $\Pi_{\mathrm{off}}$ map element-dependent orbital blocks into a shared latent set of fixed SO(3) irreps by summing over the symmetry-allowed couplings between the row and column orbital irreps. These projections are implemented as equivariant tensor contractions based on Wigner-$3j$ coefficients \cite{geiger2022e3nn}.

\textbf{Equivariant processor.}
The projected Hamiltonian features are injected into a stack of Equiformer-style transformer blocks \cite{liao2023equiformer}. At layer $k$, the node update takes the form:
\begin{equation}
\mathbf{x}_i^{(k+1)} =
\mathrm{TransBlock}^{(k)}
\left(
\mathbf{x}_i^{(k)}
+ W_{\mathrm{diag}}^{(k)} \mathbf{u}_{ii}
+ \sum_{j \neq i} W_{\mathrm{off}}^{(k)} \mathbf{u}_{ij},
\mathcal{E}
\right),
\end{equation}
where $\mathcal{E}$ denotes the geometric edge features. This expression matches the code path: diagonal Hamiltonian features are mapped to node space, off-diagonal features are linearly mapped and summed onto their destination atoms, and the result is processed by equivariant attention and feed-forward residual blocks.

\textbf{Block reconstruction.}
The model does not decode the Hamiltonian directly from every layer. Instead, the last two processor blocks feed two additional equivariant heads: a self head that accumulates node-wise latent features $\mathbf{f}_{ii}$ and a pair head that accumulates ordered pair features $\mathbf{f}_{ij}$. These latents are then expanded back to matrix blocks:
\begin{equation}
\mathbf{B}_{ii} = \mathrm{Expand}_{\mathrm{diag}}(\mathbf{f}_{ii}; \boldsymbol{\phi}_i),
\end{equation}
\begin{equation}
\mathbf{B}_{ij} = \mathrm{Expand}_{\mathrm{off}}(\mathbf{f}_{ij}; \boldsymbol{\phi}_{ij}),
\end{equation}
The expansion map follows the QHNet design~\cite{yu2023efficient}. It recombines the latent irreps into matrix entries along symmetry-allowed coupling paths defined by the row and column orbital irreps. The final prediction is symmetrized explicitly:
\begin{equation}
\widehat{\mathbf{H}}_{ii} = \mathbf{B}_{ii} + \mathbf{B}_{ii}^{\top},
\end{equation}
\begin{equation}
\widehat{\mathbf{H}}_{ij} = \mathbf{B}_{ij} + \mathbf{B}_{ji}^{\top},
\label{eq:decoder-offdiag}
\end{equation}
followed by element-specific masking. The decoder therefore enforces real symmetry at the block level while retaining a shared parameterization across different atomic species.

\textbf{Implementation Detail.}
We implement HamEvo using PyTorch~\cite{paszke2019pytorch} and e3nn~\cite{geiger2022e3nn}. In the default configuration, the radial basis has 32 channels, $\ell_{\max}=4$, and the processor contains six transformer blocks. The node hidden states use multiplicity 96 for each irrep order, and Hamiltonian block features use multiplicity 32. Fixed points are solved with Anderson acceleration during the forward pass, and implicit differentiation uses Broyden's method in the backward pass.

\textbf{Discussion.}
Two implementation choices are important for the operator view. First, the current Hamiltonian is injected at every processor block, not only at the input, so each refinement step remains explicitly state-conditioned. Second, the decoder reconstructs block matrices and symmetrizes them before masking, which matches the algebraic structure expected from the target Hamiltonian. These choices make the network closer to an iterative update rule than to a generic graph-level regressor.

\subsection{Training methodology}
\label{method:training}
The HamEvo workflow consists of two stages for constructing the base model, followed by task-specific fine-tuning for downstream applications.
Stage 1 learns single-step SCF dynamics from intermediate Hamiltonian trajectories via supervised learning~\cite{hegde2017machine, zhang2024self, neuralscf}.
Stage 2 calibrates the equilibrium fixed point using converged Hamiltonian and density-matrix supervision~\cite{wang2024infusing, li2024neural}.
Stage 3 initializes from the calibrated model and fine-tunes it on task-specific reference data, such as larger molecules or target exchange-correlation functionals.

\textbf{Stage 1: supervised learning of SCF dynamics.}
The first stage trains HamEvo to approximate an individual SCF iteration step by learning from complete convergence trajectories. The Stage 1 training data are SCF trajectories from conventional DFT calculations. For each training molecule, we run DFT calculations using the default settings described in Sec.~\ref{method:datasets}, starting from a minimal atomic orbital (MINAO) initialization, and record the Hamiltonian at each SCF iteration.
We construct the training dataset by extracting consecutive Hamiltonian pairs from these trajectories: 
\begin{equation} 
\mathcal{D}_{\text{stage1}} = \{ (\mathbf{H}_m^{(t)}, \mathbf{H}_m^{(t+1)}, \mathcal{G}_m) \mid m \in \mathcal{M}_{\text{stage1}},\ t \in \mathcal{T}_m \}
\end{equation}
where $\mathcal{M}_{\text{stage1}}$ denotes the Stage 1 molecule set, $\mathcal{G}_m = (\mathbf{Z}_m, \mathbf{R}_m)$, and $\mathcal{T}_m$ denotes the SCF iteration indices retained for training. The retained iterations are selected during dataset construction as described in Sec.~\ref{method:datasets}.

The training objective combines element-wise MAE and RMSE terms for robustness: 
\begin{multline}
\mathcal{L}_{\text{HEO}} = \mathbb{E}_{(\mathbf{H}^{(t)}, \mathbf{H}^{(t+1)}, \mathcal{G}) \sim \mathcal{D}_{\text{stage1}}} \Bigl[
\operatorname{MAE}\!\left( \mathcal{F}_\theta(\mathbf{H}^{(t)}; \mathcal{G}), \mathbf{H}^{(t+1)} \right) \\
+ \operatorname{RMSE}\!\left( \mathcal{F}_\theta(\mathbf{H}^{(t)}; \mathcal{G}), \mathbf{H}^{(t+1)} \right) \Bigr]
\end{multline} 
This objective trains HamEvo to approximate the local update dynamics of SCF iterations~\cite{zhang2024self}.

\textbf{Stage 2: physics-informed equilibrium calibration.}
While Stage 1 teaches HamEvo to approximate individual SCF steps, this approach faces a fundamental limitation: optimizing for accurate intermediate steps does not guarantee accurate equilibrium states or physical properties \cite{wang2024infusing}. 
First, iteratively applying the learned single-step function $\mathcal{F}_\theta$ can compound small prediction errors near convergence. This may lead to oscillations or divergence even when individual update steps are accurate.
Second, element-wise Hamiltonian supervision at convergence is insufficient to ensure accurate physical properties. In Kohn-Sham DFT, many ground-state observables depend on the density matrix constructed from the occupied eigenvectors~\cite{kohn1965self}. Small Hamiltonian errors that alter the occupied subspace can therefore lead to large property errors, motivating the density-matrix supervision used below~\cite{li2024neural}. Together, these considerations motivate direct fixed-point optimization with density-matrix co-supervision.

\textbf{Direct fixed-point optimization.} Rather than training only on individual update steps,
we define the model's output as the fixed point $\mathbf{H}^{\star}$ of its learned function $\mathcal{F}_\theta$, satisfying the self-consistency condition:
\begin{equation}    
\mathbf{H}^{\star} = \mathcal{F}_\theta(\mathbf{H}^{\star}; \mathcal{G})
\end{equation}
During training, we obtain $\mathbf{H}^{\star}$ by iterating $\mathcal{F}_\theta$ to convergence, then directly optimize this equilibrium state. This reduces the mismatch between one-step training and fixed-point inference by making the fixed point itself the optimization target.

Backpropagating through all fixed-point iterations would be memory-intensive. We therefore compute gradients with implicit differentiation, treating $\mathbf{H}^{\star}$ as the solution of the fixed-point equation~\cite{bai2019deep,wang2024infusing}. For a loss $\mathcal{L}(\mathbf{H}^{\star})$, the backward pass requires solving the linear system
\begin{equation}
\left(\mathbf{I} - \frac{\partial \mathcal{F}_\theta}{\partial \mathbf{H}}\bigg|_{\mathbf{H}^{\star}}\right)^{\top}\mathbf{v}
=
\frac{\partial \mathcal{L}}{\partial \mathbf{H}^{\star}}.
\end{equation}
The resulting vector $\mathbf{v}$ is then used to compute parameter gradients,
\begin{equation}
\frac{\partial \mathcal{L}}{\partial \theta}
=
\mathbf{v}^{\top}
\frac{\partial \mathcal{F}_\theta}{\partial \theta}\bigg|_{\mathbf{H}^{\star}}.
\end{equation}
In practice, we avoid forming the Jacobian explicitly and solve the linear system iteratively in the backward pass.

\textbf{Density matrix co-supervision.} To bridge the gap between element-wise Hamiltonian accuracy and property accuracy, we augment the fixed-point optimization with density matrix supervision. We construct the Stage 2 dataset from converged DFT references:

\begin{equation}
\mathcal{D}_{\text{stage2}} = \{ (\mathbf{H}^{\text{c}}_m, \mathcal{G}_m) \mid m \in \mathcal{M}_{\text{stage2}} \}
\end{equation}
where $\mathcal{M}_{\text{stage2}}$ denotes the set of molecules used in Stage 2, $\mathcal{G}_m = (\mathbf{Z}_m, \mathbf{R}_m)$, and $\mathbf{H}^{\text{c}}_m$ denotes the converged DFT Hamiltonian. For each sample, the reference density matrix $\mathbf{P}^{\text{c}}_m$ is computed online from $\mathbf{H}^{\text{c}}_m$ and the corresponding AO overlap matrix. Since ground-state properties in Kohn-Sham DFT derive from the electron density matrix $\mathbf{P}$, we augment our equilibrium loss with density matrix regularization \cite{hu2024self, li2024neural}: 

\begin{equation}
\mathcal{L} = \mathcal{L}_{\text{EQ}} + \gamma \cdot \mathcal{L}_{\text{dm}}
\end{equation}
\begin{equation}
\mathcal{L}_{\text{EQ}} = \mathbb{E}_{(\mathbf{H}^{\text{c}}, \mathcal{G}) \sim \mathcal{D}_{\text{stage2}}} \left[ \operatorname{MAE}\!\left( \mathbf{H}^{\star}, \mathbf{H}^{\text{c}} \right) + \operatorname{RMSE}\!\left( \mathbf{H}^{\star}, \mathbf{H}^{\text{c}} \right) \right]
\end{equation}
\begin{equation}
\mathcal{L}_{\text{dm}} = \mathbb{E}_{(\mathbf{H}^{\text{c}}, \mathcal{G}) \sim \mathcal{D}_{\text{stage2}}} \left[ \operatorname{MAE}\!\left( \mathbf{P}^{\star}, \mathbf{P}^{\text{c}} \right) + \operatorname{RMSE}\!\left( \mathbf{P}^{\star}, \mathbf{P}^{\text{c}} \right) \right]
\end{equation}
Here, $\mathcal{L}_{\text{EQ}}$ maintains Hamiltonian accuracy, $\mathcal{L}_{\text{dm}}$ constrains the occupied eigenspace, and the weighting parameter $\gamma$ balances these objectives.

For both the predicted fixed point $\mathbf{H}^{\star}$ and ground-truth $\mathbf{H}^\text{c}$, we compute their corresponding density matrices $\mathbf{P}^{\star}$ and $\mathbf{P}^\text{c}$ by first solving the generalized eigenvalue problem:
\begin{equation}
    \mathbf{H}\mathbf{C} = \mathbf{S}\mathbf{C}\boldsymbol{\varepsilon}
\end{equation}
where $\mathbf{S}$ is the atomic orbital overlap matrix, $\mathbf{C}$ contains the molecular orbital coefficients, and $\boldsymbol{\varepsilon}$ the orbital energies. We construct the density matrix from the $N_\text{occ} = N_\text{elec}/2$ lowest-energy orbitals:
\begin{equation}
    \mathbf{P} = 2\mathbf{C}_\text{occ}\mathbf{C}_\text{occ}^T
\end{equation}
The density matrix term directly supervises the occupied subspace that determines many ground-state observables~\cite{li2024neural}.

\textbf{Stage 3: task-specific fine-tuning.}
For downstream applications, Stage 3 uses the calibrated Stage 2 model as initialization and fine-tunes it on task-specific reference data.
The optimization follows the same fixed-point equilibrium objective as Stage 2, with the reference dataset replaced by the target large-molecule or target-functional data.
This stage is used for the large-molecule and cross-functional adaptation experiments in Sec.~\ref{sec:results_transfer} and Sec.~\ref{sec:results_functional}.

\textbf{Model Inference and Convergence.}
At inference time, HamEvo predicts the converged Hamiltonian by iterating the learned update $\mathcal{F}_\theta$ from an initial guess $\mathbf{H}^{(0)}$. Unless otherwise specified, we initialize $\mathbf{H}^{(0)}$ from the MINAO guess used in the corresponding DFT protocol~\cite{weigend2005balanced}. Anderson acceleration is used during the fixed-point iteration~\cite{anderson1965iterative}. The iteration stops when $\lVert \mathbf{H}^{(t+1)} - \mathbf{H}^{(t)} \rVert_F$ falls below the convergence tolerance, and the final iterate is returned as $\mathbf{H}^{\star}$.

\textbf{Implementation Details.}
In Stage 1, we train for 19 epochs with a batch size of 16. The learning rate starts at 1e-3 with 1,000-step warmup and follows a linear decay schedule~\cite{loshchilov2017decoupled}.
In Stage 2, we initialize the model with the pre-trained weights from Stage 1 and fine-tune for 50 epochs. The weighting factor $\gamma$ for the density matrix loss $\mathcal{L}_{\text{dm}}$ is set to 0.03, empirically chosen to balance matrix precision with spectral fidelity. For computational efficiency, the density matrix loss is introduced in the final 3,000 training steps.
To ensure numerical stability during the forward fixed-point iterations, we employ Anderson mixing with a memory size of $m = 2$ and a mixing parameter $\beta = 0.9$.
We also apply gradient clipping \cite{paszke2019pytorch} with a maximum norm of 0.05 to prevent exploding gradients, which is critical for equilibrium models. Overall, the two-stage pre-training takes approximately 264 GPU-days on A800 GPUs (40GB).

\textbf{Discussion.}
The base-model training strategy represents a form of physical curriculum learning, effectively decoupling the challenge of learning local electron interactions from the challenge of global self-consistency.
By first supervising the single-step evolution (Stage 1), we explicitly enforce the contractivity of the operator $\mathcal{F}_\theta$, ensuring that the learned dynamics are stable and physically meaningful \cite{wang2024infusing}. This pre-conditioning is crucial. Without it, direct equilibrium optimization (Stage 2) often fails to converge or collapses into trivial solutions.
Furthermore, the integration of density matrix supervision addresses a critical blind spot in traditional Hamiltonian learning: the ``spectral gap'' between matrix element accuracy and property accuracy \cite{hu2024self}.
By optimizing the density matrix $\mathbf{P}^{\star}$, we force the model to prioritize the correct ordering of eigenstates near the Fermi level, which is the primary determinant of chemical reactivity \cite{li2024neural}.
Ultimately, this methodology transforms the training process from a numerical regression task into a physics-constrained optimization, ensuring that HamEvo yields solutions that are not only mathematically convergent but also chemically accurate.

\subsection{Datasets and Computational Details}
\label{method:datasets}

In this section, we introduce the datasets and computational details employed to benchmark HamEvo. We utilize four public benchmarks (MD17~\cite{chmiela2017machine}, QM9~\cite{ramakrishnan2014quantum}, GDB17~\cite{ruddigkeit2012enumeration}, and QMugs~\cite{isert2022qmugs}), alongside one large-scale transfer learning dataset, nablaDFT~\cite{khrabrov2022nabladft}, to evaluate prediction performance across diverse chemical scales. Additionally, we specify the default DFT settings used for generating ground-truth labels and detail the specific train-test splits to ensure reproducibility.

\textbf{Default DFT Settings}. All DFT calculations in this work are performed using the GPU4PySCF-CUDA11x v1.1.0 software package~\cite{sun2020recent} at the B3LYP/def2-SVP level of theory~\cite{becke1993density, weigend2005balanced}. We employ a convergence criterion of $10^{-10}$ Hartree for SCF iterations and grid level 3 for numerical integration. Unless otherwise specified, all reported Hamiltonian matrices and derived properties are obtained under these settings.

\textbf{Evaluation Protocol}.
\phantomsection
\label{method:evaluation}
For each predicted Hamiltonian, we solve the generalized eigenvalue problem together with the corresponding reference AO overlap matrix $\mathbf{S}$ to obtain molecular-orbital energies and coefficients. Frontier orbital energies, HOMO--LUMO gaps, density matrices, dipole moments, and total energies are then computed directly from the predicted Hamiltonian through the same post-processing workflow used for the DFT references. The reported total-energy metric is the total energy error per atom. Orbital similarity is measured as the cosine similarity between AO coefficient vectors. Full metric definitions and averaging procedures are provided in Appendix~\ref{appendix:metric_define}.

\textbf{Public Benchmark}. 
The Ethanol subset from the MD17 dataset contains 25,000 training, 500 validation, and 4,500 test conformations~\cite{chmiela2017machine}. All structures are recomputed with our default DFT settings to ensure label consistency, and the full Hamiltonian evolution trajectories are retained for HamEvo training. This dataset serves to evaluate prediction performance across the conformational space of a single, flexible molecule.

Derived from the QM9 dataset~\cite{ramakrishnan2014quantum}, QH9~\cite{yu2023qh9} contains 130,831 equilibrium geometries of small molecules with up to 9 heavy atoms, comprising five elements (C, H, O, N, F). We evaluate on the QH9-stable-id task, which includes in-distribution (ID) and out-of-distribution (OOD) splits, following the split protocols from previous benchmarks~\cite{li2022deep}. All structures are recomputed with our DFT protocol, preserving evolution data where applicable.

The $\nabla^2$DFT dataset~\cite{khrabrov2024nabla2dftuniversalquantumchemistry,10.1039/D2CP03966D} is a large-scale quantum chemistry benchmark containing approximately 2 million drug-like molecules from the MOSES dataset~\cite{polykovskiy2020molecular}, with over 16 million conformations spanning eight elements (C, N, S, O, F, Cl, Br, H). We use the \emph{large} training subset (99,018 molecules, 500,552 conformations) for all experiments including baseline comparisons. Structures are originally computed at the $\omega$B97X-D/def2-SVP level using Psi4~\cite{smith2020psi4}, providing energies, forces, Hamiltonian matrices, overlap matrices, and 17 molecular properties.\footnote{We filter out Br-containing molecules from both training and test sets for basis set compatibility under our def2-SVP protocol.}

\textbf{Pre-Training Dataset Construction}.
The HamEvo model undergoes two-stage training on the following datasets, which together constitute our pre-training pipeline for building a generalizable electronic structure predictor.

The GDB17 database contains computationally enumerated drug-like molecules with up to 17 heavy atoms~\cite{ruddigkeit2012enumeration}. We construct a subset by selecting molecules composed exclusively of common organic elements from the first three periods. This subset comprises 8,446 molecules with 269,005 conformations sampled from near-equilibrium geometries during DFT-level geometry optimization trajectories. These structures are calculated with our default DFT settings, and the iterative Hamiltonian evolution data are retained for training the HEO in Stage 1.
For Stage 1 trajectory data, we retain SCF steps with $\lVert \mathbf{H}^{(t+1)}-\mathbf{H}^{(t)}\rVert_1 > 10^{-7}$ Hartree and always include the final converged step, so that the dataset contains both active update steps and fixed-point maintenance examples.

The second GDB17 subset contains 16,000 molecules selected via Butina clustering~\cite{butina1999unsupervised} based on Tanimoto similarity (threshold = 0.80) of Morgan fingerprints. For each molecule, conformations are sampled from 300~K xTB-GFN0 molecular dynamics simulations~\cite{grimme2019exploration,xtb2021methods}, yielding 76,000 geometries. Hamiltonians are calculated with default DFT protocol.

QMugs, derived from ChEMBL, provides well-curated conformations that are both diverse and low-energy, emphasizing known pharmaceutical compounds with greater structural complexity than GDB17~\cite{isert2022qmugs}. We directly use conformations from the dataset and calculate Hamiltonians with DFT. The subset contains 91,200 conformations. Together with the second GDB17 subset, this QMugs subset provides the molecular pool for the equilibrium-calibration benchmark used in Sec.~\ref{sec:results_robust_predictions}.

For this size-stratified GDB17/QMugs benchmark, molecules with 38 atoms or fewer are assigned to the training split, molecules with 39 atoms to the validation split, and molecules with 40 atoms to the test split. The QMugs property benchmark reported in Sec.~\ref{sec:results_robust_predictions} contains 9,277 conformations drawn from this 40-atom test split. Molecules larger than 40 atoms are excluded from this benchmark and are instead reserved for the large-molecule transfer experiments in Sec.~\ref{sec:results_transfer}.

\textbf{Task-Specific Evaluation Datasets}.
To rigorously evaluate HamEvo's generalization capabilities beyond standard equilibrium benchmarks, we established three specialized evaluation protocols targeting distinct regimes of chemical complexity. These tasks are designed to probe the model's ability to extrapolate along critical axes: (1) scaling to macroscopic systems significantly larger than the training distribution, (2) adapting to diverse theoretical frameworks (exchange-correlation functionals), and (3) capturing dynamic electronic effects under thermal fluctuations. The specific curation and computational details for each task are described below.

\emph{Large molecule transfer.} 
Three representative molecules are selected to evaluate size extrapolation: 4CzIPN ($\mathrm{C}_{56}\mathrm{H}_{32}\mathrm{N}_{6}$, 94 atoms), a rigid conjugated emitter; Thymopentin ($\mathrm{C}_{30}\mathrm{H}_{49}\mathrm{N}_{9}\mathrm{O}_{9}$, 97 atoms), a flexible peptide; and 3DMAC-BP-CN ($\mathrm{C}_{66}\mathrm{H}_{50}\mathrm{N}_{6}$, 122 atoms), a conjugated organic system. For each molecule, 2,000 conformations are uniformly sampled from 100~ps constant-number, volume, and temperature (NVT) molecular dynamics trajectories at 300~K (1~fs timestep, Berendsen thermostat) using xTB-GFN0, starting from xTB-optimized geometries. Hamiltonians are computed at the default DFT settings.

\emph{Cross-functional adaptation.} 
The 4CzIPN conformations above are used to evaluate functional transferability, with Hamiltonians recomputed using alternative functionals (PBE0, SCAN0, $\omega$B97X, $\omega$B97M-V) while maintaining the def2-SVP basis set. The large-scale cross-functional evaluation reuses the $\nabla^2$DFT dataset structures with target functional labels. We follow the Conformation, Structure, and Scaffold partitions distributed with the $\nabla^2$DFT benchmark, which hold out conformations, molecular structures, and molecular scaffolds, respectively.

\emph{Electron-phonon coupling.} 
Two molecules are selected: pentamantane and NAI-DMAC, starting from DFT-optimized geometries. Quantum thermal molecular dynamics (QTMD) simulations are performed using i-PI~\cite{litman2024ipi} with the GFN2-xTB~\cite{GFN2-xTB2019Bannwarth} force field at various temperatures. Each trajectory runs for 50~ps with 0.5~fs timestep, slowly heated from 10~K to the target temperature. The first 5~ps equilibration period is discarded, and conformations are sampled every 0.1~ps. Hamiltonians are computed at the B3LYP/def2-SVP level.

\textbf{Discussion.}
The dataset composition is chosen to expose HamEvo to both equilibrium and off-equilibrium molecular geometries.
Unlike traditional machine learning potentials that often prioritize equilibrium geometries (e.g., QM9~\cite{ramakrishnan2014quantum}), our pre-training pipeline includes off-equilibrium molecular dynamics trajectories and diverse low-energy conformations.
This design is motivated by the operator objective. Learning $\mathcal{F}_\theta$ requires observing how the Hamiltonian changes under SCF iterations and structural perturbations, rather than only fitting converged Hamiltonians at equilibrium geometries.
Moreover, the inclusion of the large-scale nablaDFT dataset provides a critical testbed for cross-functional transferability. By evaluating on this dataset, we demonstrate that the physical interaction rules learned from standard B3LYP data can serve as a robust foundation for adapting to more complex, computationally expensive functionals (e.g., $\omega$B97X-D) with minimal data overhead~\cite{khrabrov2022nabladft}.

\section{Data Availability}

This study uses both publicly available datasets and newly generated data, all of which are made fully accessible.

Molecular geometries were obtained from the following public sources: MD17 dataset~\cite{chmiela2017machine} (water, ethanol, malondialdehyde, uracil; accessible at \url{http://quantum-machine.org/data/schnorb_hamiltonian}); QH9 dataset~\cite{li2022deep} (\url{https://github.com/divelab/AIRS/tree/main/OpenDFT/QHBench/QH9}); and nablaDFT~\cite{khrabrov2022nabladft} 
(\url{https://github.com/AIRI-Institute/nablaDFT}). For MD17 and QH9, all electronic structure data (Hamiltonian matrices, overlap matrices, SCF evolution trajectories) were recomputed under our DFT protocol to ensure consistency. For nablaDFT, we directly use the provided Hamiltonian data.

All recomputed and newly generated datasets produced in this work, including the recomputed MD17 and QH9 with SCF convergence data, pre-training datasets (Evolution and Equilibrium), and task-specific evaluation datasets (4CzIPN, Thymopentin, 3DMAC-BP-CN, pentamantane, NAI-DMAC), are available in the HamEvo data repository on Hugging Face at \url{https://huggingface.co/datasets/ZJUSCL/hamevo-data}.

\section{Code availability}
The code used to train and evaluate the model is available on \url{https://github.com/axdfhj/HamEvo_official.git}.

\bibliography{references}

\begin{thebibliography}{10}
\expandafter\ifx\csname url\endcsname\relax
  \def\url#1{\burl{#1}}\fi
\expandafter\ifx\csname urlprefix\endcsname\relax\def\urlprefix{URL }\fi
\providecommand{\bibinfo}[2]{#2}
\providecommand{\eprint}[2][]{\url{#2}}
\providecommand{\doi}[1]{\url{https://doi.org/#1}}
\bibcommenthead

\bibitem{szabo2012modern}
\bibinfo{author}{Szabo, A.} \& \bibinfo{author}{Ostlund, N.~S.}
\newblock \emph{\bibinfo{title}{Modern quantum chemistry: introduction to advanced electronic structure theory}}  (\bibinfo{publisher}{Courier Corporation}, \bibinfo{year}{2012}).

\bibitem{hohenberg1964inhomogeneous}
\bibinfo{author}{Hohenberg, P.} \& \bibinfo{author}{Kohn, W.}
\newblock \bibinfo{title}{Inhomogeneous electron gas}.
\newblock \emph{\bibinfo{journal}{Physical review}} \textbf{\bibinfo{volume}{136}}, \bibinfo{pages}{B864} (\bibinfo{year}{1964}).

\bibitem{kohn1965self}
\bibinfo{author}{Kohn, W.} \& \bibinfo{author}{Sham, L.~J.}
\newblock \bibinfo{title}{Self-consistent equations including exchange and correlation effects}.
\newblock \emph{\bibinfo{journal}{Physical review}} \textbf{\bibinfo{volume}{140}}, \bibinfo{pages}{A1133} (\bibinfo{year}{1965}).

\bibitem{butler2018machine}
\bibinfo{author}{Butler, K.~T.}, \bibinfo{author}{Davies, D.~W.}, \bibinfo{author}{Cartwright, H.}, \bibinfo{author}{Isayev, O.} \& \bibinfo{author}{Walsh, A.}
\newblock \bibinfo{title}{Machine learning for molecular and materials science}.
\newblock \emph{\bibinfo{journal}{Nature}} \textbf{\bibinfo{volume}{559}}, \bibinfo{pages}{547--555} (\bibinfo{year}{2018}).

\bibitem{von2020exploring}
\bibinfo{author}{Ar{\'u}s-Pous, J.}, \bibinfo{author}{Awale, M.}, \bibinfo{author}{Probst, D.} \& \bibinfo{author}{Reymond, J.-L.}
\newblock \bibinfo{title}{Exploring chemical space with machine learning}.
\newblock \emph{\bibinfo{journal}{Chimia}} \textbf{\bibinfo{volume}{73}}, \bibinfo{pages}{1018--1018} (\bibinfo{year}{2019}).

\bibitem{ratcliff2017challenges}
\bibinfo{author}{Ratcliff, L.~E.} \emph{et~al.}
\newblock \bibinfo{title}{Challenges in large scale quantum mechanical calculations}.
\newblock \emph{\bibinfo{journal}{Wiley Interdisciplinary Reviews: Computational Molecular Science}} \textbf{\bibinfo{volume}{7}}, \bibinfo{pages}{e1290} (\bibinfo{year}{2017}).

\bibitem{xtb2021methods}
\bibinfo{author}{Bannwarth, C.} \emph{et~al.}
\newblock \bibinfo{title}{Extended tight-binding quantum chemistry methods}.
\newblock \emph{\bibinfo{journal}{WIREs Computational Molecular Science}} \textbf{\bibinfo{volume}{11}}, \bibinfo{pages}{e1493} (\bibinfo{year}{2021}).
\newblock \urlprefix\url{https://wires.onlinelibrary.wiley.com/doi/abs/10.1002/wcms.1493}.

\bibitem{GFN2-xTB2019Bannwarth}
\bibinfo{author}{Bannwarth, C.}, \bibinfo{author}{Ehlert, S.} \& \bibinfo{author}{Grimme, S.}
\newblock \bibinfo{title}{Gfn2-xtb—an accurate and broadly parametrized self-consistent tight-binding quantum chemical method with multipole electrostatics and density-dependent dispersion contributions}.
\newblock \emph{\bibinfo{journal}{Journal of Chemical Theory and Computation}} \textbf{\bibinfo{volume}{15}}, \bibinfo{pages}{1652--1671} (\bibinfo{year}{2019}).
\newblock \urlprefix\url{https://doi.org/10.1021/acs.jctc.8b01176}.
\newblock \bibinfo{note}{PMID: 30741547}.

\bibitem{gordon2012fragmentation}
\bibinfo{author}{Gordon, M.~S.}, \bibinfo{author}{Fedorov, D.~G.}, \bibinfo{author}{Pruitt, S.~R.} \& \bibinfo{author}{Slipchenko, L.~V.}
\newblock \bibinfo{title}{Fragmentation methods: A route to accurate calculations on large systems}.
\newblock \emph{\bibinfo{journal}{Chemical Reviews}} \textbf{\bibinfo{volume}{112}}, \bibinfo{pages}{632--672} (\bibinfo{year}{2012}).

\bibitem{Bowler_2012}
\bibinfo{author}{Bowler, D.~R.} \& \bibinfo{author}{Miyazaki, T.}
\newblock \bibinfo{title}{{$\mathcal{O}(N)$} methods in electronic structure calculations}.
\newblock \emph{\bibinfo{journal}{Reports on Progress in Physics}} \textbf{\bibinfo{volume}{75}}, \bibinfo{pages}{036503} (\bibinfo{year}{2012}).
\newblock \urlprefix\url{http://dx.doi.org/10.1088/0034-4885/75/3/036503}.

\bibitem{schutt2018schnet}
\bibinfo{author}{Sch{\"u}tt, K.~T.}, \bibinfo{author}{Sauceda, H.~E.}, \bibinfo{author}{Kindermans, P.-J.}, \bibinfo{author}{Tkatchenko, A.} \& \bibinfo{author}{M{\"u}ller, K.-R.}
\newblock \bibinfo{title}{Schnet--a deep learning architecture for molecules and materials}.
\newblock \emph{\bibinfo{journal}{The Journal of chemical physics}} \textbf{\bibinfo{volume}{148}} (\bibinfo{year}{2018}).

\bibitem{batzner2022nequip}
\bibinfo{author}{Batzner, S.} \emph{et~al.}
\newblock \bibinfo{title}{E(3)-equivariant graph neural networks for data-efficient and accurate interatomic potentials}.
\newblock \emph{\bibinfo{journal}{Nature Communications}} \textbf{\bibinfo{volume}{13}}, \bibinfo{pages}{2453} (\bibinfo{year}{2022}).

\bibitem{batatia2022mace}
\bibinfo{author}{Batatia, I.}, \bibinfo{author}{Kovacs, D.~P.}, \bibinfo{author}{Simm, G.}, \bibinfo{author}{Ortner, C.} \& \bibinfo{author}{Cs{\'a}nyi, G.}
\newblock \bibinfo{title}{Mace: Higher order equivariant message passing neural networks for fast and accurate force fields}.
\newblock \emph{\bibinfo{journal}{Advances in neural information processing systems}} \textbf{\bibinfo{volume}{35}}, \bibinfo{pages}{11423--11436} (\bibinfo{year}{2022}).

\bibitem{unke2021spooky}
\bibinfo{author}{Unke, O.~T.} \emph{et~al.}
\newblock \bibinfo{title}{Spookynet: Learning force fields with electronic degrees of freedom and nonlocal effects}.
\newblock \emph{\bibinfo{journal}{Nature communications}} \textbf{\bibinfo{volume}{12}}, \bibinfo{pages}{7273} (\bibinfo{year}{2021}).

\bibitem{Grisafi2019}
\bibinfo{author}{Grisafi, A.} \emph{et~al.}
\newblock \bibinfo{title}{Transferable machine-learning model of the electron density}.
\newblock \emph{\bibinfo{journal}{ACS Central Science}} \textbf{\bibinfo{volume}{5}}, \bibinfo{pages}{57--64} (\bibinfo{year}{2019}).
\newblock \urlprefix\url{https://doi.org/10.1021/acscentsci.8b00551}.

\bibitem{Jorgensen2022}
\bibinfo{author}{J{\o}rgensen, P.~B.} \& \bibinfo{author}{Bhowmik, A.}
\newblock \bibinfo{title}{Equivariant graph neural networks for fast electron density estimation of molecules, liquids, and solids}.
\newblock \emph{\bibinfo{journal}{npj Computational Materials}} \textbf{\bibinfo{volume}{8}}, \bibinfo{pages}{183} (\bibinfo{year}{2022}).
\newblock \urlprefix\url{https://doi.org/10.1038/s41524-022-00863-y}.

\bibitem{neuralscf}
\bibinfo{author}{Song, F.} \& \bibinfo{author}{Feng, J.}
\newblock \bibinfo{title}{Neural network self-consistent fields for density functional theory}.
\newblock \emph{\bibinfo{journal}{npj Computational Materials}}  (\bibinfo{year}{2026}).
\newblock \urlprefix\url{https://doi.org/10.1038/s41524-026-02110-0}.

\bibitem{schutt2019schnorb}
\bibinfo{author}{Sch{\"u}tt, K.~T.}, \bibinfo{author}{Gastegger, M.}, \bibinfo{author}{Tkatchenko, A.}, \bibinfo{author}{M{\"u}ller, K.-R.} \& \bibinfo{author}{Maurer, R.~J.}
\newblock \bibinfo{title}{Unifying machine learning and quantum chemistry with a deep neural network for molecular wavefunctions}.
\newblock \emph{\bibinfo{journal}{Nature communications}} \textbf{\bibinfo{volume}{10}}, \bibinfo{pages}{5024} (\bibinfo{year}{2019}).

\bibitem{unke2021phisnet}
\bibinfo{author}{Unke, O.} \emph{et~al.}
\newblock \bibinfo{title}{Se(3)-equivariant prediction of molecular wavefunctions and electronic densities}.
\newblock \emph{\bibinfo{journal}{Advances in Neural Information Processing Systems}} \textbf{\bibinfo{volume}{34}}, \bibinfo{pages}{14434--14447} (\bibinfo{year}{2021}).

\bibitem{li2022deep}
\bibinfo{author}{Li, H.} \emph{et~al.}
\newblock \bibinfo{title}{Deep-learning density functional theory hamiltonian for efficient ab initio electronic-structure calculation}.
\newblock \emph{\bibinfo{journal}{Nature Computational Science}} \textbf{\bibinfo{volume}{2}}, \bibinfo{pages}{367--377} (\bibinfo{year}{2022}).

\bibitem{yu2023efficient}
\bibinfo{author}{Yu, H.}, \bibinfo{author}{Xu, Z.}, \bibinfo{author}{Qian, X.}, \bibinfo{author}{Qian, X.} \& \bibinfo{author}{Ji, S.}
\newblock \bibinfo{title}{Efficient and equivariant graph networks for predicting quantum hamiltonian}.
\newblock \emph{\bibinfo{journal}{International Conference on Machine Learning}} \bibinfo{pages}{40412--40424} (\bibinfo{year}{2023}).

\bibitem{luo2025efficient}
\bibinfo{author}{Luo, E.} \emph{et~al.}
\newblock \bibinfo{title}{{Efficient and Scalable Density Functional Theory {H}amiltonian Prediction through Adaptive Sparsity}}.
\newblock \bibinfo{howpublished}{ICML 2025 poster} (\bibinfo{year}{2025}).
\newblock \urlprefix\url{https://icml.cc/virtual/2025/poster/45656}.

\bibitem{li2025enhancing}
\bibinfo{author}{Li, Y.} \emph{et~al.}
\newblock \bibinfo{title}{Enhancing the scalability and applicability of kohn-sham hamiltonians for molecular systems}.
\newblock \bibinfo{howpublished}{ICLR 2025 Spotlight} (\bibinfo{year}{2025}).
\newblock \urlprefix\url{https://openreview.net/forum?id=twEvvkQqPS&noteId=rc63eJkois}.

\bibitem{wang2024infusing}
\bibinfo{author}{Wang, Z.} \emph{et~al.}
\newblock \bibinfo{title}{Infusing self-consistency into density functional theory hamiltonian prediction via deep equilibrium models}.
\newblock \emph{\bibinfo{journal}{Advances in Neural Information Processing Systems}} \textbf{\bibinfo{volume}{37}}, \bibinfo{pages}{89652--89681} (\bibinfo{year}{2024}).

\bibitem{kim2025highorder}
\bibinfo{author}{Kim, S.}, \bibinfo{author}{Kim, N.}, \bibinfo{author}{Kim, D.} \& \bibinfo{author}{Ahn, S.}
\newblock \bibinfo{title}{High-order equivariant flow matching for density functional theory hamiltonian prediction}.
\newblock \bibinfo{howpublished}{NeurIPS 2025 Spotlight Poster} (\bibinfo{year}{2025}).
\newblock \urlprefix\url{https://openreview.net/forum?id=iFIjNXb0Y5}.

\bibitem{chmiela2017machine}
\bibinfo{author}{Chmiela, S.} \emph{et~al.}
\newblock \bibinfo{title}{Machine learning of accurate energy-conserving molecular force fields}.
\newblock \emph{\bibinfo{journal}{Science advances}} \textbf{\bibinfo{volume}{3}}, \bibinfo{pages}{e1603015} (\bibinfo{year}{2017}).

\bibitem{ramakrishnan2014quantum}
\bibinfo{author}{Ramakrishnan, R.}, \bibinfo{author}{Dral, P.~O.}, \bibinfo{author}{Rupp, M.} \& \bibinfo{author}{Von~Lilienfeld, O.~A.}
\newblock \bibinfo{title}{Quantum chemistry structures and properties of 134 kilo molecules}.
\newblock \emph{\bibinfo{journal}{Scientific data}} \textbf{\bibinfo{volume}{1}}, \bibinfo{pages}{1--7} (\bibinfo{year}{2014}).

\bibitem{ruddigkeit2012enumeration}
\bibinfo{author}{Ruddigkeit, L.}, \bibinfo{author}{Van~Deursen, R.}, \bibinfo{author}{Blum, L.~C.} \& \bibinfo{author}{Reymond, J.-L.}
\newblock \bibinfo{title}{Enumeration of 166 billion organic small molecules in the chemical universe database gdb-17}.
\newblock \emph{\bibinfo{journal}{Journal of chemical information and modeling}} \textbf{\bibinfo{volume}{52}}, \bibinfo{pages}{2864--2875} (\bibinfo{year}{2012}).

\bibitem{isert2022qmugs}
\bibinfo{author}{Isert, C.}, \bibinfo{author}{Atz, K.}, \bibinfo{author}{Jim{\'e}nez-Luna, J.} \& \bibinfo{author}{Schneider, G.}
\newblock \bibinfo{title}{Qmugs, quantum mechanical properties of drug-like molecules}.
\newblock \emph{\bibinfo{journal}{Scientific Data}} \textbf{\bibinfo{volume}{9}}, \bibinfo{pages}{273} (\bibinfo{year}{2022}).

\bibitem{bai2019deep}
\bibinfo{author}{Bai, S.}, \bibinfo{author}{Kolter, J.~Z.} \& \bibinfo{author}{Koltun, V.}
\newblock \bibinfo{title}{Deep equilibrium models}.
\newblock \emph{\bibinfo{journal}{Advances in neural information processing systems}} \textbf{\bibinfo{volume}{32}} (\bibinfo{year}{2019}).

\bibitem{schutt2019unifying}
\bibinfo{author}{Sch{\"u}tt, K.~T.}, \bibinfo{author}{Gastegger, M.}, \bibinfo{author}{Tkatchenko, A.}, \bibinfo{author}{M{\"u}ller, K.-R.} \& \bibinfo{author}{Maurer, R.~J.}
\newblock \bibinfo{title}{Unifying machine learning and quantum chemistry with a deep neural network for molecular wavefunctions}.
\newblock \emph{\bibinfo{journal}{Nature communications}} \textbf{\bibinfo{volume}{10}}, \bibinfo{pages}{5024} (\bibinfo{year}{2019}).

\bibitem{unke2021se}
\bibinfo{author}{Unke, O.} \emph{et~al.}
\newblock \bibinfo{title}{Se (3)-equivariant prediction of molecular wavefunctions and electronic densities}.
\newblock \emph{\bibinfo{journal}{Advances in Neural Information Processing Systems}} \textbf{\bibinfo{volume}{34}}, \bibinfo{pages}{14434--14447} (\bibinfo{year}{2021}).

\bibitem{sun2020recent}
\bibinfo{author}{Sun, Q.} \emph{et~al.}
\newblock \bibinfo{title}{Recent developments in the pyscf program package}.
\newblock \emph{\bibinfo{journal}{The Journal of chemical physics}} \textbf{\bibinfo{volume}{153}} (\bibinfo{year}{2020}).

\bibitem{kummel2008orbital}
\bibinfo{author}{K{\"u}mmel, S.} \& \bibinfo{author}{Kronik, L.}
\newblock \bibinfo{title}{Orbital-dependent density functionals: Theory and applications}.
\newblock \emph{\bibinfo{journal}{Reviews of Modern Physics}} \textbf{\bibinfo{volume}{80}}, \bibinfo{pages}{3} (\bibinfo{year}{2008}).

\bibitem{gong2023general}
\bibinfo{author}{Gong, X.} \emph{et~al.}
\newblock \bibinfo{title}{General framework for e (3)-equivariant neural network representation of density functional theory hamiltonian}.
\newblock \emph{\bibinfo{journal}{Nature Communications}} \textbf{\bibinfo{volume}{14}}, \bibinfo{pages}{2848} (\bibinfo{year}{2023}).

\bibitem{merchant2023scaling}
\bibinfo{author}{Merchant, A.} \emph{et~al.}
\newblock \bibinfo{title}{Scaling deep learning for materials discovery}.
\newblock \emph{\bibinfo{journal}{Nature}} \textbf{\bibinfo{volume}{624}}, \bibinfo{pages}{80--85} (\bibinfo{year}{2023}).

\bibitem{wang2024deeph}
\bibinfo{author}{Wang, Y.} \emph{et~al.}
\newblock \bibinfo{title}{Deeph-2: Enhancing deep-learning electronic structure via an equivariant local-coordinate transformer}.
\newblock \emph{\bibinfo{journal}{arXiv preprint arXiv:2401.17015}}  (\bibinfo{year}{2024}).

\bibitem{li2024neural}
\bibinfo{author}{Li, Y.} \emph{et~al.}
\newblock \bibinfo{title}{Neural-network density functional theory based on variational energy minimization}.
\newblock \emph{\bibinfo{journal}{Physical Review Letters}} \textbf{\bibinfo{volume}{133}}, \bibinfo{pages}{076401} (\bibinfo{year}{2024}).

\bibitem{hu2024self}
\bibinfo{author}{Hu, G.} \emph{et~al.}
\newblock \bibinfo{title}{Self-consistent validation for machine learning electronic structure}.
\newblock \emph{\bibinfo{journal}{arXiv preprint arXiv:2402.10186}}  (\bibinfo{year}{2024}).

\bibitem{zhang2024self}
\bibinfo{author}{Zhang, H.} \emph{et~al.}
\newblock \bibinfo{title}{Self-consistency training for density-functional-theory hamiltonian prediction}.
\newblock \emph{\bibinfo{journal}{arXiv preprint arXiv:2403.09560}}  (\bibinfo{year}{2024}).

\bibitem{weigend2005balanced}
\bibinfo{author}{Weigend, F.} \& \bibinfo{author}{Ahlrichs, R.}
\newblock \bibinfo{title}{Balanced basis sets of split valence, triple zeta valence and quadruple zeta valence quality for h to rn: Design and assessment of accuracy}.
\newblock \emph{\bibinfo{journal}{Physical Chemistry Chemical Physics}} \textbf{\bibinfo{volume}{7}}, \bibinfo{pages}{3297--3305} (\bibinfo{year}{2005}).

\bibitem{becke1993density}
\bibinfo{author}{Becke, A.~D.}
\newblock \bibinfo{title}{Density-functional thermochemistry. iii. the role of exact exchange}.
\newblock \emph{\bibinfo{journal}{The Journal of Chemical Physics}} \textbf{\bibinfo{volume}{98}}, \bibinfo{pages}{5648--5652} (\bibinfo{year}{1993}).

\bibitem{chai2008long}
\bibinfo{author}{Chai, J.-D.} \& \bibinfo{author}{Head-Gordon, M.}
\newblock \bibinfo{title}{Long-range corrected hybrid density functionals with damped atom--atom dispersion corrections}.
\newblock \emph{\bibinfo{journal}{Physical Chemistry Chemical Physics}} \textbf{\bibinfo{volume}{10}}, \bibinfo{pages}{6615--6620} (\bibinfo{year}{2008}).

\bibitem{khrabrov2022nabladft}
\bibinfo{author}{Khrabrov, K.} \emph{et~al.}
\newblock \bibinfo{title}{nabladft: Large-scale conformational energy and hamiltonian prediction benchmark and dataset}.
\newblock \emph{\bibinfo{journal}{Physical Chemistry Chemical Physics}} \textbf{\bibinfo{volume}{24}}, \bibinfo{pages}{25853--25863} (\bibinfo{year}{2022}).

\bibitem{kovacs2023mace}
\bibinfo{author}{Kovács, D.~P.} \emph{et~al.}
\newblock \bibinfo{title}{Mace-off23: Transferable machine learning force fields for organic molecules} (\bibinfo{year}{2023}).
\newblock \urlprefix\url{https://arxiv.org/abs/2312.15211}.
\newblock \bibinfo{eprint}{{\href{https://arxiv.org/abs/2312.15211}{{arXiv:2312.15211}}}}.

\bibitem{mathiasen2024reducing}
\bibinfo{author}{Mathiasen, A.} \emph{et~al.}
\newblock \bibinfo{title}{Reducing the cost of quantum chemical data by backpropagating through density functional theory}.
\newblock \emph{\bibinfo{journal}{arXiv preprint arXiv:2402.04030}}  (\bibinfo{year}{2024}).

\bibitem{roothaan1951new}
\bibinfo{author}{Roothaan, C. C.~J.}
\newblock \bibinfo{title}{New developments in molecular orbital theory}.
\newblock \emph{\bibinfo{journal}{Reviews of modern physics}} \textbf{\bibinfo{volume}{23}}, \bibinfo{pages}{69} (\bibinfo{year}{1951}).

\bibitem{geiger2022e3nn}
\bibinfo{author}{Geiger, M.} \& \bibinfo{author}{Smidt, T.}
\newblock \bibinfo{title}{e3nn: Euclidean neural networks}.
\newblock \emph{\bibinfo{journal}{arXiv preprint arXiv:2207.09453}}  (\bibinfo{year}{2022}).

\bibitem{liao2023equiformer}
\bibinfo{author}{Liao, Y.-L.} \& \bibinfo{author}{Smidt, T.}
\newblock \bibinfo{title}{Equiformer: Equivariant graph attention transformer for 3d atomistic graphs}.
\newblock \emph{\bibinfo{journal}{arXiv preprint arXiv:2206.11990}}  (\bibinfo{year}{2022}).

\bibitem{paszke2019pytorch}
\bibinfo{author}{Paszke, A.} \emph{et~al.}
\newblock \bibinfo{title}{Pytorch: An imperative style, high-performance deep learning library}.
\newblock \emph{\bibinfo{journal}{Advances in neural information processing systems}} \textbf{\bibinfo{volume}{32}} (\bibinfo{year}{2019}).

\bibitem{hegde2017machine}
\bibinfo{author}{Hegde, G.} \& \bibinfo{author}{Bowen, R.~C.}
\newblock \bibinfo{title}{Machine-learned approximations to density functional theory hamiltonians}.
\newblock \emph{\bibinfo{journal}{Scientific reports}} \textbf{\bibinfo{volume}{7}}, \bibinfo{pages}{42669} (\bibinfo{year}{2017}).

\bibitem{anderson1965iterative}
\bibinfo{author}{Anderson, D.~G.}
\newblock \bibinfo{title}{Iterative procedures for nonlinear integral equations}.
\newblock \emph{\bibinfo{journal}{Journal of the ACM (JACM)}} \textbf{\bibinfo{volume}{12}}, \bibinfo{pages}{547--560} (\bibinfo{year}{1965}).

\bibitem{loshchilov2017decoupled}
\bibinfo{author}{Loshchilov, I.} \& \bibinfo{author}{Hutter, F.}
\newblock \bibinfo{title}{Decoupled weight decay regularization}.
\newblock \emph{\bibinfo{journal}{arXiv preprint arXiv:1711.05101}}  (\bibinfo{year}{2017}).

\bibitem{yu2023qh9}
\bibinfo{author}{Yu, H.} \emph{et~al.}
\newblock \bibinfo{title}{{QH9}: A quantum hamiltonian prediction benchmark for qm9 molecules}.
\newblock \emph{\bibinfo{journal}{arXiv Preprint, arXiv:2306.09549}}  (\bibinfo{year}{2023}).

\bibitem{khrabrov2024nabla2dftuniversalquantumchemistry}
\bibinfo{author}{Khrabrov, K.} \emph{et~al.}
\newblock \bibinfo{title}{$\nabla^2$dft: A universal quantum chemistry dataset of drug-like molecules and a benchmark for neural network potentials}.
\newblock \bibinfo{howpublished}{NeurIPS 2024 Track Datasets and Benchmarks Poster} (\bibinfo{year}{2024}).
\newblock \urlprefix\url{https://openreview.net/forum?id=ElUrNM9U8c}.
\newblock \bibinfo{eprint}{{\href{https://arxiv.org/abs/2406.14347}{{arXiv:2406.14347}}}}.

\bibitem{10.1039/D2CP03966D}
\bibinfo{author}{Khrabrov, K.} \emph{et~al.}
\newblock \bibinfo{title}{nabladft: Large-scale conformational energy and hamiltonian prediction benchmark and dataset}.
\newblock \emph{\bibinfo{journal}{Phys. Chem. Chem. Phys.}} \textbf{\bibinfo{volume}{24}}, \bibinfo{pages}{25853--25863} (\bibinfo{year}{2022}).
\newblock \urlprefix\url{http://dx.doi.org/10.1039/D2CP03966D}.

\bibitem{polykovskiy2020molecular}
\bibinfo{author}{Polykovskiy, D.} \emph{et~al.}
\newblock \bibinfo{title}{Molecular sets (moses): a benchmarking platform for molecular generation models}.
\newblock \emph{\bibinfo{journal}{Frontiers in pharmacology}} \textbf{\bibinfo{volume}{11}}, \bibinfo{pages}{565644} (\bibinfo{year}{2020}).

\bibitem{smith2020psi4}
\bibinfo{author}{Smith, D.~G.} \emph{et~al.}
\newblock \bibinfo{title}{Psi4 1.4: Open-source software for high-throughput quantum chemistry}.
\newblock \emph{\bibinfo{journal}{The Journal of chemical physics}} \textbf{\bibinfo{volume}{152}} (\bibinfo{year}{2020}).

\bibitem{butina1999unsupervised}
\bibinfo{author}{Butina, D.}
\newblock \bibinfo{title}{Unsupervised data base clustering based on daylight's fingerprint and tanimoto similarity: A fast and automated way to cluster small and large data sets}.
\newblock \emph{\bibinfo{journal}{Journal of Chemical Information and Computer Sciences}} \textbf{\bibinfo{volume}{39}}, \bibinfo{pages}{747--750} (\bibinfo{year}{1999}).

\bibitem{grimme2019exploration}
\bibinfo{author}{Grimme, S.}
\newblock \bibinfo{title}{Exploration of chemical compound, conformer, and reaction space with meta-dynamics simulations based on tight-binding quantum chemical calculations}.
\newblock \emph{\bibinfo{journal}{Journal of chemical theory and computation}} \textbf{\bibinfo{volume}{15}}, \bibinfo{pages}{2847--2862} (\bibinfo{year}{2019}).

\bibitem{litman2024ipi}
\bibinfo{author}{Litman, Y.} \emph{et~al.}
\newblock \bibinfo{title}{i-pi 3.0: a flexible and efficient framework for advanced atomistic simulations}.
\newblock \emph{\bibinfo{journal}{J. Chem. Phys.}} \textbf{\bibinfo{volume}{161}}, \bibinfo{pages}{062505} (\bibinfo{year}{2024}).

\end{thebibliography}

\begin{figure*}[t!]
\centering
\includegraphics[width=0.98\textwidth]{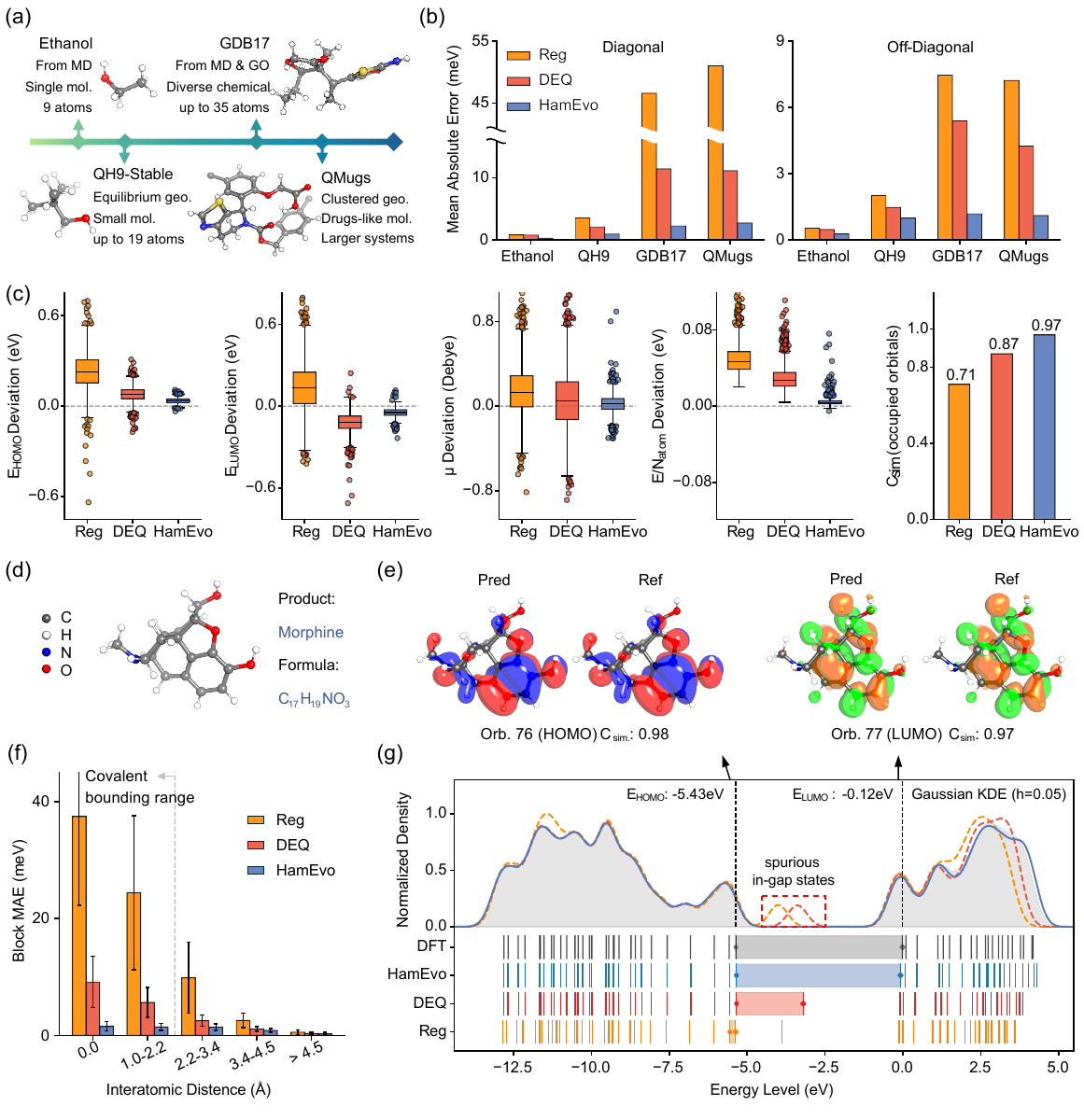}
% \vskip -0.1in
\caption{Accurate predictions across chemical space. \textbf{(a)} Evaluation datasets ranging from simple ethanol (MD17) to complex drug-like molecules (QMugs). \textbf{(b)} Comparison of Hamiltonian MAEs. HamEvo (blue) consistently outperforms the baselines. \textbf{(c)} Accuracy comparison of derived electronic observables on QMugs, including errors in $E_{\mathrm{HOMO}}$, $E_{\mathrm{LUMO}}$, and $\mu$, together with occupied-orbital similarity ${C_{\text{sim}}}$. \textbf{(d)} Molecular structure of Morphine, used for the case study in (e--g). \textbf{(e)} Visualization of frontier orbitals. Predicted HOMO and LUMO isosurfaces (Pred) closely match the DFT references (Ref). \textbf{(f)} Distance-binned Hamiltonian block MAEs for Morphine. HamEvo achieves the largest improvement in the short-range covalent-bonding region. \textbf{(g)} Orbital-energy distribution for Morphine. The HamEvo spectrum follows the DFT reference and preserves the HOMO--LUMO gap, whereas the baselines introduce spurious in-gap states.}
\label{fig:general_results}
\vskip -0.2in
\end{figure*}

\clearpage
\appendix
\renewcommand{\thesection}{A} 

\section*{Appendix}

% \subsection{Architecture Detail}
% \label{appendix:full_architecture_details}

% xxx

% \textbf{Pretraining Dataset.}

% \textbf{Optimization Details.}

\subsection{Definition of Evaluation Metrics}
\label{appendix:metric_define}

\textbf{Hamiltonian MAE ($\text{MAE}_{\mathbf{H}}$).}
To assess the numerical accuracy of the predicted electronic structure at the matrix level, we define the Mean Absolute Error of the Hamiltonian ($\text{MAE}_{\mathbf{H}}$). This metric measures the element-wise discrepancy between the predicted Hamiltonian matrix $\mathbf{H}^{\text{pred}}$ and the reference DFT Hamiltonian $\mathbf{H}^{\text{DFT}}$. To avoid confusion with the atom-pair block notation $\mathbf{H}_{ij}$ used in the main text, we use $\mu,\nu$ here to denote AO basis-function indices of the full Hamiltonian matrix:

\begin{equation}
\label{appendix:definition_maeh}
\text{MAE}_{\mathbf{H}} = \frac{1}{N_{\text{basis}}^2} \sum_{\mu=1}^{N_{\text{basis}}} \sum_{\nu=1}^{N_{\text{basis}}} \left| (\mathbf{H}^{\text{pred}})_{\mu\nu} - (\mathbf{H}^{\text{DFT}})_{\mu\nu} \right|
\end{equation}

where $N_{\text{basis}}$ denotes the total number of AO basis functions, and $(\mathbf{H})_{\mu\nu}$ denotes a scalar matrix element in the AO basis. This global average provides a comprehensive measure of the model's ability to reconstruct the entire Hamiltonian operator across all AO basis-function pairs.

\textbf{Property Deviations ($\Delta$).}
For each test conformation $m$, we define the signed deviation of a derived property as the predicted value minus the DFT reference value. The frontier-orbital energy deviations are
\begin{equation}
\Delta E_{\mathrm{HOMO},m} = \varepsilon_{\mathrm{HOMO},m}^{\text{pred}} - \varepsilon_{\mathrm{HOMO},m}^{\text{DFT}},
\end{equation}
\begin{equation}
\Delta E_{\mathrm{LUMO},m} = \varepsilon_{\mathrm{LUMO},m}^{\text{pred}} - \varepsilon_{\mathrm{LUMO},m}^{\text{DFT}}.
\end{equation}
For the dipole moment, we compute the deviation in the magnitude of the dipole vector,
\begin{equation}
\Delta \mu_m = \left\| \boldsymbol{\mu}_m^{\text{pred}} \right\| - \left\| \boldsymbol{\mu}_m^{\text{DFT}} \right\|.
\end{equation}
For the total energy metric reported in Fig.~\ref{fig:general_results}c, we use the signed total-energy error per atom,
\begin{equation}
\Delta (E/N_{\mathrm{atom}})_m =
\frac{E_{\mathrm{total},m}^{\text{pred}} - E_{\mathrm{total},m}^{\text{DFT}}}{N_{\mathrm{atom},m}}.
\end{equation}
The plotted property errors are mean absolute deviations over the corresponding test set,
\begin{equation}
\operatorname{MAE}(q) = \frac{1}{N_{\mathrm{test}}} \sum_{m=1}^{N_{\mathrm{test}}} \left| \Delta q_m \right|,
\end{equation}
where $q$ denotes any of the derived properties above.

\textbf{Orbital Similarity (${C_{\text{sim}}}$).}
To evaluate the spatial fidelity of the predicted molecular orbitals, we define the orbital similarity metric, ${C_{\text{sim}}}$. This metric quantifies the overlap between the predicted molecular orbitals and the ground-truth orbitals obtained from Density Functional Theory (DFT).

In the context of the LCAO approximation, each molecular orbital is represented by a coefficient vector $\mathbf{C}_i$ in the AO basis set. For a given molecular system, the similarity is calculated by averaging the cosine similarity between the predicted coefficient vectors $\mathbf{C}_i^{\text{pred}}$ and the reference DFT vectors $\mathbf{C}_i^{\text{DFT}}$ over all occupied orbitals ($N_{\text{occ}}$):

\begin{equation}
\label{appendix:definition_csim}
{C_{\text{sim}}} = \frac{1}{N_{\text{occ}}} \sum_{i=1}^{N_{\text{occ}}} \frac{\left| \mathbf{C}_i^{\text{pred}} \cdot \mathbf{C}_i^{\text{DFT}} \right|}{\left\| \mathbf{C}_i^{\text{pred}} \right\| \cdot \left\| \mathbf{C}_i^{\text{DFT}} \right\|}
\end{equation}

where $\|\cdot\|$ denotes the $L_2$ norm. The use of the absolute value in the numerator ensures that the metric is invariant to the arbitrary phase (sign) of the wavefunctions. A value of ${C_{\text{sim}}}$ close to 1 indicates that the model has high fidelity in capturing the wavefunction shapes, nodal structures, and spatial distributions, thereby preserving the essential symmetry and chemical nature (e.g., $\sigma$ and $\pi$ character) of the molecular orbitals.

\begin{table*}[htbp]
\centering
\caption{
    Derived-property prediction accuracy on the QMugs test set.
    Errors are reported as MAEs for $E_{\mathrm{HOMO}}$ (eV), $E_{\mathrm{LUMO}}$ (eV), and $\mu$ (Debye).
    Orbital similarity is reported as the mean ${C_{\text{sim}}}$ over occupied orbitals.
    \textbf{Bold} indicates the best performance.
}
\label{tab:derived_property_accuracy}
\renewcommand{\arraystretch}{1.15}
\setlength{\tabcolsep}{6pt}
\begin{tabular}{lcccc}
\toprule
\textbf{Method} & $\mathbf{E_{\mathrm{HOMO}}}$ & $\mathbf{E_{\mathrm{LUMO}}}$ & $\boldsymbol{\mu}$ & $\mathbf{C_{\text{sim}}}$ \\
\midrule
QHMet  & 0.525 & 3.401 & 9.741 & 0.534 \\
DEQ    & 0.082 & 0.129 & 0.250 & 0.865 \\
HamEvo & \textbf{0.036} & \textbf{0.053} & \textbf{0.098} & \textbf{0.974} \\
\bottomrule
\end{tabular}
\end{table*}

\iffalse
\textbf{Root-Mean-Square Deviation (RMSD).}
To quantify the geometric variations between molecular conformations, particularly for structures sampled from Molecular Dynamics (MD) trajectories, we utilize the Root-Mean-Square Deviation (RMSD). This metric evaluates the average atomic displacement between two sets of coordinates. 

The calculation process consists of two primary stages:
(1) \textit{Superposition}: Before computing the deviation, the two conformations must be optimally aligned to eliminate the influence of global translation and rotation. This involves shifting the center of mass of both structures to the origin and applying the Kabsch algorithm to find the optimal rotation matrix that minimizes the RMSD.
(2) \textit{Calculation}: Following the optimal superposition, the RMSD (measured in \AA) is calculated as:
\begin{equation}
\label{appendix:definition_rmsd}
\text{RMSD} = \sqrt{\frac{1}{N} \sum_{i=1}^{N} \left[ (x_i - x'_i)^2 + (y_i - y'_i)^2 + (z_i - z'_i)^2 \right]}
\end{equation}
where $N$ denotes the number of atoms, and $(x_i, y_i, z_i)$ and $(x'_i, y'_i, z'_i)$ represent the Cartesian coordinates of the $i$-th atom in the two conformations, respectively. A small RMSD value indicates a high degree of structural similarity between the datasets, ensuring that the sampled conformations represent physically reasonable variations around the equilibrium geometry.
\fi

% \subsection{Cross Function Finetuning Detail}
% \label{appendix:cross_function_finetune_detail}

% Data for cross function generalization.
\begin{table*}[htbp]
\centering
\caption{
    Hamiltonian prediction accuracy (MAE, in meV) across diverse molecular benchmarks. 
    Results are reported for diagonal, off-diagonal, and total matrix elements. 
    For the QH9 benchmark, ID and OOD denote the in-distribution and out-of-distribution splits, respectively.
    \textbf{Bold} indicates the best performance.
}
\label{tab:hamiltonian_mae}
\renewcommand{\arraystretch}{1.15}
\setlength{\tabcolsep}{5pt}
\begin{tabular}{ll cccccc}
\toprule
\multirow{2}{*}{\textbf{Method}} & \multirow{2}{*}{\textbf{Metric}} 
    & \textbf{MD17} & \textbf{QH9} & \textbf{QH9} & \textbf{QH9} & \multirow{2}{*}{\textbf{GDB17}} & \multirow{2}{*}{\textbf{QMugs}} \\
    & & \textbf{Ethanol} & \textbf{ID} & \textbf{OOD} & \textbf{Geo-mol} & & \\
\midrule
\multirow{3}{*}{QHNet}
    & Diag      & 0.85  & 3.51  & 3.04  & 11.30 & 46.60  & 51.03 \\
    & Off-diag  & 0.49  & 2.01  & 1.90  & 4.18  & 7.46  & 7.22  \\
    & Total     & 0.57  & 2.11  & 1.96  & 4.73  & 8.58  & 8.61  \\
\cmidrule(lr){1-8}
\multirow{3}{*}{DEQHNet}
    & Diag      & 0.77  & 2.08  & 2.20  & 5.74  & 11.40 & 11.10 \\
    & Off-diag  & 0.47  & 1.47  & 1.41  & 2.64  & 5.39  & 4.24  \\
    & Total     & 0.51  & 1.51  & 1.45  & 2.88  & 5.61  & 4.46  \\
\cmidrule(lr){1-8}
\multirow{3}{*}{\textbf{HamEvo}}
    & Diag      & \textbf{0.25}  & \textbf{0.92}  & \textbf{1.10}  & \textbf{4.14}  & \textbf{2.24}  & \textbf{2.70}  \\
    & Off-diag  & \textbf{0.26}  & \textbf{0.98}  & \textbf{0.97}  & \textbf{2.14}  & \textbf{1.18}  & \textbf{1.10}  \\
    & Total     & \textbf{0.26}  & \textbf{0.98}  & \textbf{0.98}  & \textbf{2.29}  & \textbf{1.22}  & \textbf{1.15}  \\
\bottomrule
\end{tabular}
\end{table*}

\end{document}